\documentclass[
amsmath, amssymb, 
aps, pra,
reprint, twocolumn,
superscriptaddress,
longbibliography,
floatfix,
showkeys
]{revtex4-1}

\usepackage{graphicx}
\usepackage{csquotes}
\usepackage{multirow}
\usepackage[urlcolor=blue, colorlinks=true,citecolor=blue]{hyperref}

\usepackage{color}
\usepackage{nicefrac}

\definecolor{orange}{RGB}{255,127,0}

\def\avg#1{\mathinner{\langle{#1}\rangle}}
\def\bra#1{\ensuremath{\mathinner{\langle{#1}|}}}
\def\ket#1{\ensuremath{\mathinner{|{#1}\rangle}}}

\newcommand{\braket}[2]{\langle #1|#2\rangle}

\DeclareMathOperator*{\argmin}{arg\,min}

\input{Qcircuit}

\begin{document}

\title{Low depth mechanisms for quantum optimization}

\author{Jarrod R. McClean}
	\email{Corresponding author: jmcclean@google.com}
	\affiliation{Google Research, 340 Main Street, Venice, CA 90291, USA}
\author{Matthew P. Harrigan}
	\affiliation{Google Research, 340 Main Street, Venice, CA 90291, USA}
\author{Masoud Mohseni}
	\affiliation{Google Research, 340 Main Street, Venice, CA 90291, USA}
\author{Nicholas C. Rubin}
	\affiliation{Google Research, 340 Main Street, Venice, CA 90291, USA}
\author{Zhang Jiang}
	\affiliation{Google Research, 340 Main Street, Venice, CA 90291, USA}	
\author{Sergio Boixo}
	\affiliation{Google Research, 340 Main Street, Venice, CA 90291, USA}
\author{Vadim N. Smelyanskiy}
	\affiliation{Google Research, 340 Main Street, Venice, CA 90291, USA}
\author{Ryan Babbush}	
	\affiliation{Google Research, 340 Main Street, Venice, CA 90291, USA}
\author{Hartmut Neven}
	\affiliation{Google Research, 340 Main Street, Venice, CA 90291, USA}
\date{\today}     

\begin{abstract}
One of the major application areas of interest for both near-term and fault-tolerant quantum computers is the optimization of classical objective functions.  In this work, we develop intuitive constructions for a large class of these algorithms based on connections to simple dynamics of quantum systems, quantum walks, and classical continuous relaxations.  We focus on developing a language and tools connected with kinetic energy on a graph for understanding the physical mechanisms of success and failure to guide algorithmic improvement.  This physical language, in combination with uniqueness results related to unitarity, allow us to identify some potential pitfalls from kinetic energy fundamentally opposing the goal of optimization.  This is connected to effects from wavefunction confinement, phase randomization, and shadow defects lurking in the objective far away from the ideal solution.  As an example, we explore the surprising deficiency of many quantum methods in solving uncoupled spin problems and how this is both predictive of performance on some more complex systems while immediately suggesting simple resolutions.  Further examination of canonical problems like the Hamming ramp or bush of implications show that entanglement can be strictly detrimental to performance results from the underlying mechanism of solution in approaches like QAOA.  Kinetic energy and graph Laplacian perspectives provide new insights to common initialization and optimal solutions in QAOA as well as new methods for more effective layerwise training.  Connections to classical methods of continuous extensions, homotopy methods, and iterated rounding suggest new directions for research in quantum optimization.  Throughout, we unveil many pitfalls and mechanisms in quantum optimization using a physical perspective, which aim to spur the development of novel quantum optimization algorithms and refinements.
\end{abstract}

\maketitle

\section{Introduction}
Optimization is a topic so broad reaching and powerful in its applications that the idea that it could be possibly accelerated by quantum computers has attracted incredible attention, regardless of  the origins or the underlying mechanisms for speedup.  Issues in optimization are at the core of problems ranging from applied problems in logistics, financial trading, and machine learning to problems of theoretical interest in the study of spin glasses.  However an intuitive understanding of exactly why and how the coherence and entanglement of a quantum computation would assist in classical optimization problems can be elusive.  The goal of this work is to highlight and improve mechanisms that occur even for low depth quantum circuits, to enable better engineering of quantum optimization.

Grover's algorithm, quantum walks and their adaptations were among the first to suggest that quantum computers may provide an advantage for problems in optimization~\cite{Grover:1996,Durr1996Quantum,szegedy2004quantum}.  It was shown that for an unstructured optimization, quantum computers could provide a quadratic advantage.  In practice however, the mechanism by which Grover proceeds uses coherent transport between a known initial state, and marked final state (or set of states), producing few or no approximate solutions in between.  It seems unlikely that for a real problem on $1000$ bits, one can afford to wait a time on the order of $\sim 2^{500}$ for the solution, without any meaningful improvement beyond random guessing before that.  While this argument is a bit of a straw-man, it's meant to illustrate that blindly applying quantum techniques to optimization often does not yield a practical algorithm.  Moreover, there may be some development in quantum oracles required to even admit the desired quadratic speedup.  In a naive example, for NP-Hard spin glass problems built from $k$-local terms, queries to the Hamiltonian in classical algorithms reuse work from previous steps by computing only differences on spin configurations, which can reduce the complexity of the query by a factor of $n$ compared to a naive quantum oracle query which often by construction checks all terms simultaneously. Hence if the direct use of Grover is to make any impact in optimization problems, it is likely to be in a narrow window of problems with fewer than $100$ bits input and those for which no better algorithm than random guessing is known.  

Using a more modern interpretation of Grover search and amplitude amplification has lead to asymptotic speedups over structured classical algorithms like annealing and branch-and-bound~\cite{cerf2000nested,durr2006quantum,mandra2016faster,ambainis2019quantum,montanaro2020quantum}, but the overheads for general cases remain challenging when compiled all the way to fault tolerant gate sequences~\cite{sanders2020compilation}. This is exacerbated by the fact that real use cases have shown that it is likely the case that improved solutions are most needed on problems $>1000$ bits, where classical heuristics break down~\cite{campbell2019applying,heim2015quantum}. While connections to searching by continuous time quantum walks have provided better bases to understand some of these results~\cite{Childs2004spatial}, the mechanism leaves much to be desired for understanding where quantum computers may provide optimization wins beyond polynomial.

The study of quantum annealing has also provided insight into areas where quantum computers may assist with optimization~\cite{kadowaki1998quantum,farhi2001quantum,johnson2011quantum,boixo2014evidence,denchev2016computational}.  The works in quantum annealing and adiabatic quantum optimization offer concrete  mechanisms that could lead to quantum advantage. The physical mechanisms are often attributed to faster coherent collective tunneling through energy landscape barriers, and that difficult problems are associated with phase transitions and closures of the eigenvalue gap along the adiabatic path. However, the potential computational power of quantum annealers over classical techniques is yet not well understood as they could suffer from decoherence effects, finite control precision, and sparse connectivity. Moreover, physical quantum annealers have low dimensional graphs and typically encode strongly disordered problems and thus could suffer from exponentially small gaps due to Griffith's singularities \cite{Mohseni18}. Some of these
limitations could be avoided by inhomogeneous quantum annealing
algorithms motivated by Kibble-Zurek mechanism ~\cite{Mohseni18,Nishimori2018} or hybrid quantum-assisted Monte Carlo techniques ~\cite{Mohseni_patent}. However, similar physics-inspired techniques could also be implemented in classical methods combining certain cluster update strategies over a backbone algorithm from the MCMC family (e.g., simulated annealing~\cite{Kirkpatrick671} or parallel tempering (PT) algorithms~\cite{PTreview}).  The cluster updates could 
include Swendsen-Wang-Wolf cluster updates~\cite{Swendsen_Wang87,Wolf89}, Hodayer
moves~\cite{Houdayer2001}, or Hamze-Freitas-Selbey~\cite{HF04,Selby14,Hen_2017}, and approximate tensor network contractions \cite{Rams2018}. %
However, these classical approaches either break down for frustrated systems ~\cite{Wolf89}, or percolate for $D>2$ \cite{Houdayer2001} or assume random tree-like subgraphs ~\cite{HF04,Selby14,Hen_2017}, so in general their generalization to Monte Carlo based cluster updates to higher dimensional frustrated systems are not yet fully explored. Additionally, recent advances in machine learning could lead to new techniques for discovery of such collective latent variables for higher dimensional spin-glass systems that could encode discrete optimization problems \cite{hartnett2020}.

From a physical perspective, quantum tunneling through barriers, despite a quadratic advantage in the many-body case~\cite{smelyanskiy2018non} over classical diffusion as in Langevin methods for global optimization~\cite{gidas1985global}, scales exponentially in the width of the barrier~\cite{denchev2016computational,smelyanskiy2018non,kechedzhi2018efficient}.  Moreover in the single-body case, a host of results demonstrate that quantum annealing, which are similar to Langevin methods, can be efficiently simulated on classical computers with quantum Monte Carlo~\cite{crosson2016simulated,jiang2017scaling,jiang2017path}. Hence, even if one is able to take advantage of these mechanisms, one may find themselves in a Grover-like situation, where practically cannot wait $\sim 2^w$, where $w$ is a generic many-body barrier width sometimes scaling as $O(n)$ for glassy problems, to see any improvement of solution.  This is especially true for near-term and even early fault-tolerant devices with limited quantum resources.

For problems without any structure, quadratic speedups are the best one can hope for, and indeed this leads to a steep overhead for practical problems that can look insurmountable for near-term devices~\cite{sanders2020compilation}.  However, this opens the questions of if there are structures in classical optimization problems that quantum computers are uniquely posed to take advantage of that could lead to super-polynomial speedups, and if these structures are present in classical optimization problems of interest.  While there is some fear that the presence of structure in a problem might be equally exploitable by classical algorithms, we recall that without specific and somewhat sensitive structure, Shor's factoring algorithm~\cite{shor1999polynomial} would also not succeed.  Indeed there is conjecture that for the majority of problems, in a measure-theoretic sense, beyond polynomial speedups are not possible without structure~\cite{aaronson2009need}.  Hence, understanding such structures is key to understanding the potential of quantum computers for optimization.

In lieu of the identification of such possible structures, researchers have exploited numerical and theoretical studies of collections of well-known optimization problems.  The hope is certainly that if quantum computers could take advantage of some structure, it may be present in these problems, and these effects might be observed in the form of a computational speedup.  However, even if one takes the viewpoint that such structures are present in some instances of a particular problem, if that problem is difficult enough, there may be an equal or greater number of adversarial instances that cancel the desired benefit when averaged over a sufficient number of examples.  Given the sensitivity of known quantum algorithms to precise phase coherence, it seems that for a problem with a wide enough degree of expressibility, (e.g. NP-Complete problems), practical solvers must allow quantum algorithms to be used on instances they are best at, without compromising their performance for instances they are not well suited for.  Moreover, at small system sizes, even some difficult spin-glass problems display considerable local structure that classical solvers can take advantage of, while structure-less quantum solvers (e.g. a naive application of Grover's algorithm or in some cases quantum annealing) are needlessly hamstrung, making some blind explorations stilted comparisons at best.

The desire to explore the impact of quantum computers on real problems without the corresponding hardware to perform algorithms of exponential depth has led to the development of a number of heuristics, including the variational quantum eigensolver and the quantum approximate optimization algorithm~\cite{Peruzzo:2014,McClean:2016Theory,OMalley:2016,Farhi:2014,nannicini2019performance}.  The applicability of these algorithms to optimization problems draws intuition in a high-depth limit from quantum annealing and adiabatic quantum optimization~\cite{kadowaki1998quantum,farhi2000quantum}.  However, these algorithms are operated in the low-depth regime, and despite some recent strides in this direction~\cite{zhou2018quantum,hastings2019classical,bravyi2019obstacles,brandao2018fixed}, understanding which optimization problems they might excel at remains challenging.  Without a mechanistic understanding of how optimization can operate in these circumstances, it is difficult if not impossible to improve the algorithms performances substantially.

\begin{figure}[t!]
\centering
\includegraphics[width=8cm]{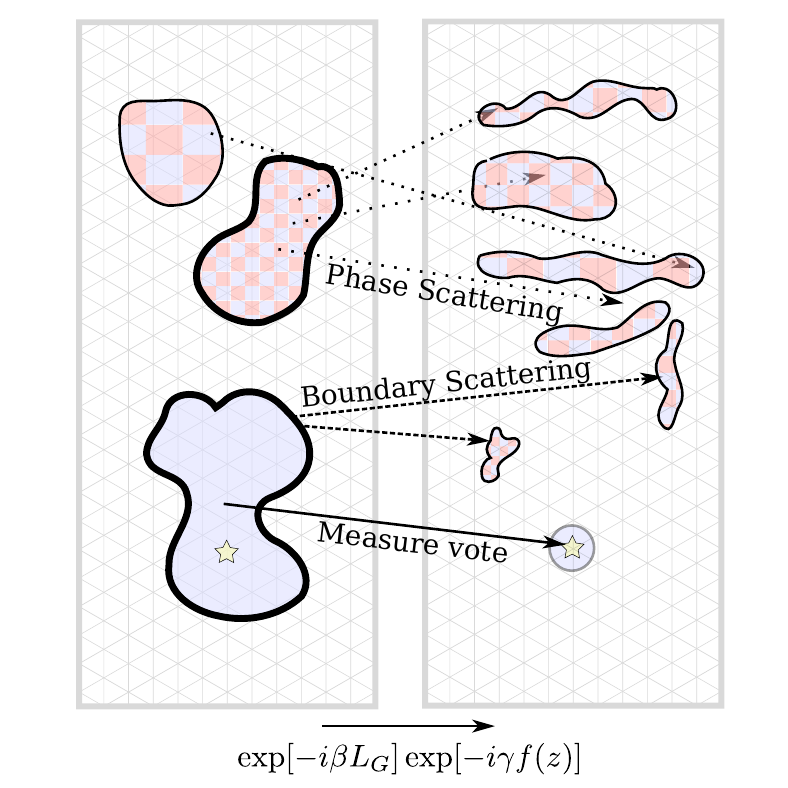}
    \caption{Cartoon of mechanisms in low depth optimization discussed here.  On the left and right a particular wavefunction is depicted where the vertices are computational basis states, $\ket{z}$, and colored regions represent wavefunction support where colors depict relative phases between sites, i.e. the same color regions have the same associated phase.  The transition from left to right depicts evolution under a low-depth optimization primitive like QAOA under a potential $f(z)$ and graph Laplacian $L_G$.  The different processes depicted represent different mechanisms that may happen at once or at different times in an algorithm.  As one progresses in an algorithm, the domain of support for possible solutions contracts and relative phase information accumulates inside domains of support.  It is found that subsequent steps have dynamics generated by scattering from the boundary of the domain and randomized relative internal phases.  These processes can detract from primary mechanisms of solution, such as concentrating density on a solution where most domain potential points agree, or a measure vote.
    \label{fig:OverviewCartoon}}
\end{figure}

In this work, we use the low-depth regime as a lens for gaining intuition that can lead to real improvement in quantum algorithms for optimization.  Some of this intuition is cartooned in Fig \ref{fig:OverviewCartoon}.  We identify structure in classical optimization problems that quantum computers excel at, rather than hope such structure will appear without knowing its form.  In doing so, we note some oversights that allow for immediate improvement in almost all quantum algorithms for quantum optimization.  Through connections to the language of continuous time quantum walks and understanding quantum optimization as fortuitous coherence on the underlying graph, we find a new perspective in understanding these approaches.  Although we will focus on short-depth realizations to maximize understanding, these tools turn out to be powerful for better understanding of how quantum optimizations based on physical dynamics proceed in general.

\section{Summary of Topics and Outline}
In order to better organize this work for the time conscious reader, we attempt to outline the general topics of discussion and conclusions from each section.  In Section \ref{sec:OptAndWalks} we begin with a review of optimization and connections to the language of quantum walks.  We highlight the formulation of several popular approaches such as quantum annealing, adiabatic optimization, and the quantum approximate optimization algorithm (QAOA) in a way that allows them to be simply connected and generalized in the language of graph Laplacians.

Following this, in Section \ref{sec:ExactSols} we take common cases from the literature, and using properties of unitarity demonstrate a number of exact results on problems solvable in extremely short depth.  We show that the Grover driver (or complete graph Laplacian) represents an extreme of too many connections, preventing any structure in a problem from being exploited in a meaningful way under these constructions in contrast with other possible choices for the graph.  We go on to show that existing methods, such as quantum annealing and a single step QAOA can fail to solve uncoupled spin problems in surprising ways.  While some of these results are known in the literature, our exposition aims to increase clarity.  These constructions lead us to analogies with problems of scale like ill-conditioning in gradient descent classically, and methods for preconditioning problems to avoid these difficulties.

This brings us to a more general connection to continuous extensions in Section \ref{sec:ContinuousExts} and the relationship between VQE and the classical multi-linear extension.  In particular, we see that many methods for variational circuits used on discrete optimization problems share commonalities with classical techniques often used in approximation methods and belief propagation.  These similarities motivate the introduction of a technique from classical approximation methods, iterated rounding, which addresses problems of scale, quantum-classical readout challenges, and underlying challenges related to symmetry.  A connection is also made to the classical theory of homotopy methods for global optimization, which bridge the quantum adiabatic method with classical approaches.

Using this information we identify a mechanism of solution in short-depth algorithms like QAOA, and construct a mean-field variant to exhibit its properties in Section \ref{sec:MeasureVote}.  This mechanism operates by an effective majority vote of the potential surface.  In particular, several of the canonical problems studied, such as the spike problem on a Hamming ramp and the bush of implications suggest that while a single step of QAOA both solves these problems efficiently, and is entangled, in these cases that entanglement is strictly detrimental to performance.  Modifications to avoid these problems based on the flexibility of the graph Laplacian are introduced based on mechanistic insights from the mean-field model.

This global consensus model reveals a particular class of failures that can occur within quantum optimization methods we term ``shadow defects'' in Section \ref{sec:ShadowDefects}.  In particular, energetic defects distant in Hamming weight can have considerable influence on the ability of variational methods to effectively refine their solutions due to the global nature of evolutions.  We explore simple examples where this is the case and the impact on the efficiency of algorithms.  Our conjecture is that this phenomenon is an underlying mechanism related to concentration of measure observed in QAOA landscapes for random problem instances  \cite{brandao2018fixed,Verdon2019metalearning}.  Modifying the objective function, for example using a Gibbs-like objective~\cite{li2019quantum},  allows one to avoid some of these issues, but more scalable solutions require cutting space into informed, smaller problems.

This brings us to the crucial underlying element of kinetic energy in quantum optimization in Section \ref{sec:KineticEnergy}.  Kinetic energy from the confinement of a quantum optimization based on a real Hamiltonian presents a force fundamentally opposed to the goals of classical optimization.  Insights from kinetic energy allow us to understand why so many optimal paths in QAOA resemble smooth deformations, why initial guess paths from adiabatic schedules are often crucial for success, and why layerwise greedy optimizations are so challenging.  We offer techniques for circumventing these challenges, such as weighting the kinetic energy while training layerwise or searching for new variational components.

The insights related to kinetic energy motivates the development of coherent graph cutting in Section \ref{sec:CoherentCutting}.  This technique leverages the power of coherent quantum algorithms to show how it might be possible to introduce additional prior information into a classical optimization problem in a way that does not damage performance.

In Section \ref{sec:Numerics} we investigate numerically the impact of these issues and basic techniques of remediation on common models like ferromagnets and MaxCut. In analogy to non-interacting systems, simple rescaling can destroy the performance of some algorithms on systems like a 2D ferromagnet.  We also show substantial depth is required to reach a classical baseline from a simple classical continuous extension, in part due to its natural ability to adapt to scale. Finally in Section\ref{sec:ImprovingLowDepth} we tie together our conclusions to make general recommendations for the improvement of low-depth quantum optimization and close with thoughts and outlooks.

\section{Optimization and quantum walks}
\label{sec:OptAndWalks}
We begin by reviewing and establishing the notation for both quantum optimization and quantum walks as will be relevant to our narrative.  The goal in optimization is to find the value $z^*$ such that $f(z^*)$ is a maximum or minimum of the value of the objective function, $f$.  For the most part, we will restrict our discussion to classical functions $f(z)$ where the domain of $z$ is bitstrings of length $n$, due to the simplicity to map the problem to $n$ qubits. Throughout this work we will use the notation $X_i$, $Y_i$, and $Z_i$ to refer to the Pauli operators acting on qubit $i$. While there is some interest in finding local minima of problems for practical reasons, the discussion in quantum optimization often focuses on the more challenging task of finding global optima, and this can be concisely restated as
\begin{align}
    z^*= \argmin_z f(z).
\end{align}

In some cases one is interested in the typical quality of solutions a method produces, which can be quantified by the approximation ratio.  As a matter of clarification, the approximation ratio in classical discrete optimization and approximation methods has been defined differently than the approximation in reference to quantum optimization and QAOA in particular.  We choose to take the convention common in the quantum literature where an approximation ratio of $1$ is perfect, $0$ is failure, and we seek to maximize the approximation ratio, which is opposite to the convention taken in the classical optimization algorithm literature.  Hence for this choice of convention we use throughout, the approximation ratio for minimization of a solution $z$ is defined to be
\begin{align}
  R = \mathbb{E}\left[ \frac{f_\text{max} - f(z)}{f_\text{max} - f(z^*)} \right] 
\end{align}
where $f_\text{max}$ is the maximum value of $f$ and the approximation ratio is evaluated over the distribution of solutions a method produces and/or a distribution of instances for a problem.

Quantum annealing~\cite{kadowaki1998quantum} works to solve this optimization problem through connection with the adiabatic principle by evolving under Hamiltonians of the form
\begin{align}
    H = s(t) \sum_z f(z) \ket{z} \bra{z} - (1-s(t)) \sum_i X_i 
\end{align}
where one begins in the initial state $\ket{+}^{\otimes n}$ and continuously varies the schedule function from $s(0)=0$ to $s(1)=1$.   If this evolution is slow enough, in a sense which can be made precise~\cite{albash2018adiabatic}, the evolution stays in the ground state of the joint Hamiltonian, and produces the optimal bitstring $z^*$ with high probability.  We note that there is sometimes a subtle distinction in connotation between quantum annealing and the optimization through the quantum adiabatic principle in that quantum annealers built from analog components sometimes operate without the ability to exactly give $s(0)=0$, and hence approximate this path by making the coefficient of the transverse field $(- \sum_i X_i)$ large in comparison to the magnitude of $f(z)$ in initial stages of the algorithm.

One can use a similar construction for functions with only one marked state, that is $f(z) = 0$ for all but $f(z^*)=-1$ to formulate the Hamiltonian analog of Grover's algorithm~\cite{farhi1998analog}, which traditionally proceeds via the Hamiltonian
\begin{align}
    H = s(t) \sum_z f(z) \ket{z} \bra{z} - (1-s(t)) P_+
\end{align}
where $P_+ = \ket{+}\bra{+}^{\otimes n}$, which we've written in a way as to be structurally similar to the previous equation.  

At a glance, the two Hamiltonians appear quite different, and indeed in many ways they are.  However, in the literature of continuous-time quantum walks~\cite{Childs2004spatial,Novo2015}, it has been noted that there is a common structure that underlies the two, and it is this structure we will run with to develop intuition for quantum optimization of classical problems.  In particular, both algorithms start from the state $\ket{\psi_i}=\ket{+}^{\otimes n}$ and proceed via a Hamiltonian of the form
\begin{align}
    H = s(t) \sum_z f(z) \ket{z} \bra{z} + (1-s(t)) L_G
\end{align}
where $L_G$ is the graph Laplacian of a graph $G$, where the nodes are computational bitstrings or basis states.  Indeed, it turns out that for a wide family of graphs $G$, the above algorithms would produce the ground state of $f(z)$, albeit with computational scalings that depend on the specifics of the graph.  However, to date, the specifics of the graph have been under utilized in understanding and accelerating these algorithms.  Let us briefly recall what we need to know about graph Laplacians in order to unify the conceptual pictures.

At a conceptual level, the graph Laplacian is the real-space Laplacian used in quantum mechanics for the kinetic energy, generalized to generic and discrete graphs.  In particular for optimization problems we can consider them as evolving under a potential and kinetic energy defined by
\begin{align}
    T &= L_G \\
    V &= \sum_z f(z) \ket{z} \bra{z}
\end{align}
Hence, as in the Euclidean case (e.g. a 1D set of points connected in a line), it is related to the kinetic energy of a system, and evolution under this operator generates the dynamics of the system in position (or computational bases state) space.  While there are a few formal definitions in the literature of the graph Laplacian, we will take $L_G$, of a graph $G$ for an undirected, $k$-regular graph as 
\begin{align}
    L_G = D_G - A_G
\end{align}
where $D_G$ and $A_G$ are the $N \times N=2^n \times 2^n$ degree and adjacency matrices of the graph $G$.  $D_G$ is a diagonal matrix with entries for each node indicating the number of connections to that node, or degree, and $A_G$ with entries $[A_{G}]_{ij} = [A_{G}]_{ji} = 1$ if nodes $i$ and $j$ are adjacent, and $0$ otherwise.  For $k$-regular graphs, the degree matrix $D_G$ is proportional to the identity, and hence for quantum evolution it may be disregarded as imparting a global phase.  

Graph Laplacians have the important property that for connected graphs, there is a unique ground state with eigenvalue $0$, which corresponds to $\ket{+}^{\otimes n}$.  Conceptually, this is the minimum kinetic energy state for all connected graphs, and hence any application of the adiabatic principle for finding the ground state, starts with the state $\ket{+}^{\otimes n}$.  This also implies that any connected graph $G$ could in principle enable a meaningful quantum annealing schedule, so this begs the questions of why one would choose one graph over the other.  The study of continuous time quantum walks informs this question in the generic sense, but its also clear that the precise coherences will depend on the details of the matching between the node values (potential) and the underlying graph.  One could imagine that a good match between the graph and the potential may yield good performance, or an open-gap, while a poor one could result in a closing gap in the language of adiabatic quantum computation.  It is these types of correspondences we examine, but we will not restrict ourselves to the adiabatic regime.

In the above examples, we saw the two most common choices of bit-string graphs in quantum computing, namely the hypercube graph, $-\sum_i X_i$, which connects each bitstring to all neighbors one bitflip away, and the complete graph $P_+$, which connects all bit strings to all other bitstrings.  These two graphs are special cases of a type of graph we will find particularly compelling to analyze, which are so-called ``fast-forwardable graphs''.  In slight contrast to recent definitions of the fast-forwardable graphs~\cite{apers2018quantum}, we mean that evolution under the graph Laplacian, $e^{-i L_G t}$, can be done with a straightforward circuit implementation scaling at worst as $O(\log t)$ in the number of gates.  This is the case for any graph Laplacian that has a simple quantum diagonalization circuit.  For example, the Hadamard transform on all qubits, which diagonalizes both $P_+$ and sums over tensor products of $X$ terms. This also includes the euclidean families of graphs (e.g. a discretized 1D line or 2D plane) that may be diagonalized through quantum Fourier transforms, as has been noted in early works on quantum walks~\cite{Childs2004spatial}. It also includes graphs defined in a fermionic picture~\cite{wang2018quantum,wang2020,chapman2020characterization}.

The connection of Laplacians to kinetic energy $T \propto L_G$ with the current conventions will be important for developing understanding in this work.  In classical random walks or searches over the bitstrings and function $f(z)$, the fact that walkers in any given instance of time are localized to single  bit-strings is relatively unimportant.  However, in quantum evolutions, the global graph of the problem imparts a kinetic energy that can generate evolution unrelated to the problem $f(z)$ from simply confining the wavefunction to a smaller support of bitstrings.  This force acts against us in quantum optimization in a fundamental way, and can be influenced by the phase as well, with the unique $0$ kinetic energy state being the uniform phase, uniform weight state $\ket{+}^{\otimes n}$. Hence if one cannot use phase coherence to their advantage, the growing random quantum diffusion imparted from the kinetic energy operator is likely to act as a hindrance rather than an aid~\cite{nandkishore2015many}.  

The combination of results from quantum optimization with the limited resources available today has led to the development of heuristics with primitives inspired by these algorithms.  Most notable among them are the use of variational quantum eigensolver (VQE)~\cite{Peruzzo:2014} and the quantum approximate optimization algorithm (QAOA)~\cite{Farhi:2014}, sometimes also referred to as the quantum alternating operator ansatz~\cite{hadfield2019quantum}.  These approaches both use an outer-loop variational principle to determine the parameters based on a secondary variational objective to be updated.  This variational objective, we denote as $C(f(z))$ and in most cases is taken to be the expected value with respect to a wavefunction, or
\begin{align}
    C_m(f(z)) = \langle f(z) \rangle
\end{align}
where the subscript $m$ denotes expected value, however we will explore the importance of alternative $C(f(z))$'s in this work, including non-trivial additions of kinetic energy for reasons that will become apparent.

In practice, the VQE is a general construction for using parameterized circuits to find ground states, which can include quantum or classical cases, and the QAOA is a variational construction designed to include coherent queries to the objective function.  Although it is not equivalent for all parameter choices, it is sometimes characterized as a parameterized version of a low-depth Trotter expansion of the adiabatic algorithm specialized for classical objective functions.  It is often repeated for a number of repetitions $p$.  We note that by convention, the QAOA ansatz is defined with a negative convention with respect to the graph Laplacian we use here, so to maintain consistency in the optimal parameters with the literature, we use $\bar L_G = -L_G$ when appropriate.  When cast in the conventions we consider here, has the form
\begin{align}
    \min_{\beta_i, \gamma_i} \bra{\psi} f(z) \ket{\psi}; \ \ket{\psi} = \prod_i^p e^{-i \beta_i \bar L_G} e^{-i \gamma_i f(z)} \ket{+}^{\otimes n}
\end{align}
Broadly however, our results will not be specific to QAOA or VQE, but we will be able to examine the single step process at $p=1$ for considerable insight when viewed in the setting as a continuous walk over the bit string graph.

\section{Exact solutions} \label{sec:ExactSols}
Many classical optimization methods proceed by the iterative solution of a sequence of exact sub-problems or approximations thereof that materially predict the performance of methods or provide mechanistic suggestions for improvement.  An apt analogy that we draw is to that of first order optimization methods like gradient descent, which are plagued by issues related to condition numbers, and are often repaired by techniques that take note of such shortcomings such as  quasi-Newton or preconditioned methods~\cite{nocedal2006numerical}.  We aim to do the same here for low-depth quantum optimization problems by showcasing a subset of exactly solvable problems and some surprising failures, such as the inability of a traditional quantum annealing or QAOA construction to solve non-interacting spin problems exactly at $p=1$.  While some of the results here are known in more general contexts, we attempt to highlight when this is the case while focusing on a simplified presentation.  For example, while some of our results focus on the $p=1$ case, where general proofs of classical algorithms are known~\cite{hastings2019classical}, this will support our goal of understanding what happens within a differential step of a quantum optimization algorithm more generally.  In the quest for clarity, we focus on defining exact solutions to mean perfect fidelity with a ground state.  While this is restrictive and may not always track precisely with approximate results, we believe the resulting conclusions are more readily translated to strategies for improvement in all cases.

\subsection{Implications of unitarity}

While the objective of some algorithms is to maximize the approximation ratio, a sufficient condition that will be easier to analyze is obtaining the exact solution of an optimization problem, which here we take to be unique.  We denote the exact solution as $\ket{z^*}$, a computational bit string, and our goal is to understand what problems can be solved in a single step, or those for which quantum optimizations are exceptionally well suited.  We worry about corresponding classical difficulty later.  For this problem, in the standard setup we take the standard $0$ kinetic energy state $\ket{+}^{\otimes n}$.  This means exactly solving the problem is equivalent to
\begin{align}
    e^{-i \gamma f(z)} \ket{+}^{\otimes n} = e^{i \beta \bar L_G} \ket{z^*} \label{Eq:ExactConstraint}
\end{align}
where we have moved the action of the Laplacian to act on the bitstring for convenience.  From here, we may make a few simple observations that will be important.  First, for an exact equivalence to exist, the action of $e^{i \beta \bar L_G}$ be able to map a single bitstring to a state which is a uniform weight superposition over all bitstrings, although the phases may be allowed to vary.  This follows from the initial state and diagonal property of the operator $f(z)$.  This fact already greatly constrains both $L_G$ and the values of $\beta$ that are allowed as we will see below, often independent of the state $\ket{z^*}$.  Hence, if the graph $G$ is given, this often immediately implies the allowed values of $\beta$, which in turn dictates that only a small family (if not only one) potential $f(z)$ is exactly solvable up to branches of the complex logarithm.

Perhaps even more strongly, when one considers the property of unitarity, if $L_G$ implies the allowed values of $\beta$, and the problem has a solution $\ket{z^*}$, then there exists only one potential which this can exactly solve, and it is determined by the graph rather than the problem Hamiltonian.  This implies our ability to solve problems is about the interplay of the potential $f(z)$ with the graph $G$ rather than simply the problem objective $f(z)$ by itself.

Another point of consequence when considering unitarity is that even for a long sequence of $\beta_i, \gamma_i$ for a given graph $G$ and potential $f(z)$, there is no fixed $\beta_i$, $\gamma_i$, for example in a learning rate schedule that decreases $\beta_i$ and $\gamma_i$ in accordance with an adiabatic schedule for a large number of repetitions, that can solve the problem from an arbitrary initial state.  That is, given some $U$ for which $U\ket{+}^{\otimes n} = \ket{z^*}$ , there is no other state $\ket{\psi}$ such that $U\ket{\psi} = \ket{z^*}$ by unitarity.  This is seen simply from the properties of unitarity, which preserve inner products $\bra{\psi} U^\dagger U \ket{+}^{\otimes n} = \braket{\psi}{+}^{\otimes n}$.  This implies the algorithm is not globally convergent in the numerical analysis sense in that it cannot even converge to a local minimum (as defined by the graph $G$ and $f(z)$) from an arbitrary initial state, even for the simplest of potentials as in a typical stochastic gradient descent.  This appears to be a problem for any unitary quantum algorithm where the unitary is not made to depend on the initial state or extended to include non-unitary actions through ancilla.  With these general concerns in mind, we turn to more specific graphs.

\subsection{Grover and the complete graph}
We begin with the complete bitstring graph, associated with $L_G = -P_+ = -\ket{+}\bra{+}^{\otimes n}$, as is commonly used in Grover's algorithm or other common quantum walk variants of search for a marked bitstring.  Expanding the general constraint in Eq. \ref{Eq:ExactConstraint} for this graph Laplacian we find
\begin{align}
    e^{-i f(z) \gamma} \ket{+}^{\otimes n} &= e^{i \beta P_+} \ket{z^*} \notag \\
    &= \left[ I + (e^{i \beta} - 1 ) P_+ \right] \ket{z^*} \notag \\
    &= \ket{z^*} + (e^{i \beta} - 1 ) \braket{+}{z^*} \ket{+}^{\otimes n} \notag \\
    &= \ket{z^*} + \frac{1}{2^{n/2}} (e^{i \beta} - 1)\ket{+}^{\otimes n}
\end{align}
where we have only used the fact that $\ket{z^*}$ is a computational bitstring.  If we recall that we require the ability to map to a uniformly weighted state for an exact solution to be possible, we find that the most uniform weighted state is given by $\beta = \pi$, and in this case, we find that the maximum possible change in overlap in some sense scales as $\sim 1/2^{(n/2)}$.  Hence the complete graph solves exactly no potentials in one step, and makes Grover-like progress at best for any potentials, regardless of structure that might exist.  We focus here on the case of 1 step to aid in what we can say exactly and gain understanding about the relationship between the potential and the graph Laplacian.  In this sense, the complete graph is associated with a structureless algorithm in that this limit on progress had no dependence on the form of $f(z)$.  It also makes clear the statement that more connections on the bit string graph are not always better.  In a complete graph there is no ``barrier'' in a strict sense between any initial state and a target state, however the unused connections act as an effective, entropic barrier that waste useful exploration of a normalized state.  The complete graph is the extreme example of this, but we will see other graphs suffer from this phenomenon in a weaker way.  Hence, it is useful and necessary to exploit structure to weaken or cut unused connections. This is at the heart of structured probabilistic models or graphical models in classical machine learning community that can encode the structure of underlying data/problem by removing the unnecessarily direct links between variables \cite{goodfellow2016deep}.    

\subsection{Transverse field and hypercube graph}
We now turn to the hypercube graph commonly used in quantum annealing, QAOA, quantum adiabatic algorithms, and other quantum walk implementations.  This graph is associated with $L_G = -\sum_i X_i$ and is the graph with each bitstring connected to its $n$ neighbors by flipping each qubit individually.  Algebraic proofs of the limitations of this graph have been well explored in the context of QAOA~\cite{streif2019comparison,bapat2018bang}, but we aim to present these in a way that supports our coming exposition. This type of graph is sometimes called a Kronecker sum graph due to the separable form.  This fact is also suggestive of the type of potentials it can solve exactly, which we will see are separable over qubits.  Considering the constraint in Eq. \ref{Eq:ExactConstraint}, but for a single qubit we find
\begin{align}
    e^{-i \gamma f(z)} \ket{+} &= e^{i \beta X} \ket{z^*} \\
    &= (\cos \beta I + i \sin \beta X) \ket{z^*}
\end{align}
It is straightforward to find that only two values of $\beta$ satisfy the constraint of preparing a state with equal weights on all qubits, for either $z^* \in \{0, 1\}$, which are $\beta=\{\pi/4, 3\pi/4 \}$.  

However, the exact sign and state actually do depend on the value of $z^*$ in this case, and that dictates $\beta=\pi/4$.  Note however, that in some cases, the exact ground state is inconsistent with the ground state of the one-body terms.  For example, if there is a higher-body term present that dominates the determination of the ground state.  In this case, the adjustment of $\beta$ to $3\pi /4$ allows the potential orient the contributions of the single body terms correctly in order to agree with conflicting higher body terms.  This example provides mechanistic insight behind variational adjustment of $\beta$ beyond setting it to $\pi/4$.  Moreover, so long as $\beta > 0$, it is easy to stay in a regime where properties for exact adiabatic are retained.  This also suggests however, that an individual qubit $\beta$ can be significantly more powerful in the presence of conflicting terms.  Note that our use of conflicting terms throughout is consistent with the definition of non-frustration free Hamiltonians in quantum computer science where the ground state of the total Hamiltonian is not the ground state of each of its constituent parts, however to avoid confusion with geometric frustration in physical systems, we will prefer the term conflicting.

The use of the $\beta=\pi/4$ immediately yields discrete exactly solvable potentials up to branches of the complex log
\begin{align}
    - \gamma f(z) = \frac{\pi}{2}w = \frac{\pi}{4} (I + Z)
\end{align}
where $w$ is also the Hamming weight of the qubit, or Hamming distance from $0$.  The ability to adjust $\gamma$ in conjunction with redundancy implies that one can solve exactly any one qubit potential of the form $\alpha(I \pm Z)$, where $\alpha \in \mathbb{R}$ and the shift by the identity may be included or not for convenience.  As the graph Laplacian is a Kronecker sum, this must be satisfied individually for each qubit, which may be re-written as
\begin{align}
    \bigotimes_j (e^{-i \gamma f(z_j)} \ket{+}) &= \bigotimes_j (e^{i \beta X} \ket{z^*_j})
\end{align}
where $z_j$ and $z_j^*$ refer to the $j$'th bit and this form shows the separability.  From the previous set of results, we see that the only set of solvable potentials are some of the form $\alpha \sum_i (I \pm Z_i)$, where we specifically note that $\alpha$ is fixed across all spins.  This is, up to sign and unimportant shifts, the Hamming distance symmetric potential which has been shown a number of times to be solvable in one step of the QAOA algorithm~\cite{streif2019comparison,bapat2018bang}.

It is worth discussing in slightly more generality the freedom that is imparted by the branches of the complex log in the derived expressions.  That is, for $f(z)$ which is solvable on any bitstring one may add multiples of $2\pi / \gamma$ and exactly the same operator will result.  However, since $e^{-i \gamma f(z)}$ is the same operator, and the target ground state $\ket{z^*}$ has remained the same, only additions which are compatible with the original ground state allow the problem to remain exactly solvable.  Hence, only having the principal solvable potential will result in the same answer.  In the event that it alters it, we lose the property of exact solution.  In the case that it leaves it the same, we may solve this potential exactly as well.  Unfortunately, this class of potentials, due to being dictated entirely by the 1-body terms, will also be trivially solvable by the classical mean-field method we introduce later.  In order to take advantage of some quantum properties one must either pass through an intermediate entangled state dictating $p>1$ for unique bitstring solutions, or end with an entangled state.

\subsection{Odd $k$-spin ferromagnets and branches}
A recent work by Wauters and collaborators~\cite{wauters2020polynomial} showed that QAOA can prepare ground states of $k$-spin quantum Ferromagnets efficiently, and moreover as a special case there are some classical configurations (without transverse fields) that are amenable to exact $p=1$ solutions. We will show how this result fits into the construction and intuition we have built here.  In particular, classical $k$-spin ferromagnet Hamiltonians have the form
\begin{align}
    f(z) = -(\sum_i^n Z_i)^k.
\end{align}
It was shown by algebraic means, that in the case that both $k$ and $n$ are odd, a $p=1$ QAOA forms an exact solution at all system sizes for $\beta=\pi/4, \gamma=\pi/4$.  As evolution under the hypergraph Laplacian cannot generate or destroy entanglement, and the final state is unentangled, we conclude that for all of these systems, evolution under the potential can also generate no entanglement.  This is suggestive of the fact that a principle potential of uncoupled spins is likely at the source of this success.  To position this in our language here, let's consider the smallest case of $k=3$, $n=3$.  In this case, we can expand the potential to find
\begin{align}
    f(z) &= -(Z_1 + Z_2 + Z_3)^3 \notag \\
    &= -7(Z_1 + Z_2 + Z_3) - 6(Z_1 Z_2 Z_3)
\end{align}
up to shifts by the identity.  At a glance, this seems to contain the non-trivial term $Z_1 Z_2 Z_3$ and other weight factors.  However, if we rearrange the Hamiltonian into the form
\begin{align}
    f(z) = &-(Z_1 + Z_2 + Z_3) \notag \\
    &- 6(Z_1 + Z_2 + Z_3 + Z_1 Z_2 Z_3)
\end{align}
we find that the second term is equal to $-24(\ket{000}\bra{000} - \ket{111}\bra{111})$, which when multiplied by $\gamma=\pi/4$, is exactly an integer multiple of $2\pi$ on basis states, imparting no effect on the state or change in evolution under the potential operator.  Hence, the entire solution is determined by the first term of uncoupled spins that happens to not be disrupted by the second.

\subsection{Degenerate ground states}
So far we have considered the case of having a unique solution to the optimization problem, labeled $\ket{z^*}$.  When considering the hypercube graph, this had the implication of solving primarily uncoupled spin problems of fixed magnitude.  Another way of seeing this result, is to see that as the evolution operator $e^{-i \beta \bar L_G}$ is separable and can introduce no entanglement, if one starts and finishes in a separable state, the potential cannot introduce entanglement either, immediately ruling out generic interactions between qubits in the potential beyond those that introduce trivial phase factors.  To circumvent this and explore the implications of quantum entanglement, we now explore cases with degenerate solutions containing entanglement.  This section will focus on a specific case of the more general result shown for limitations in $\mathbb{Z}_2$ symmetric Hamiltonians~\cite{bravyi2019obstacles}, but will introduce some additional thoughts on the modification of the Laplacian. We note that this is of interest for the approximation capabilities of these algorithms as well, if the overlap of a unique solution is greater with an entangled state than any separable one achievable within a given circuit.

To begin we start with the simplest such potential, a pair potential of the form
\begin{align}
    f(z) = -Z_1 Z_2.
\end{align}
From a classical perspective, this has two, equally good solutions, either $z=00$ or $z=11$.  In a quantum perspective, the ground state of this diagonal Hamiltonian is degenerate, and any state within the degenerate subspace $\text{span}\{\ket{00}, \ket{11}\}$ is allowed.  We have seen from the previous section and entanglement arguments that if we make our goal to produce either then $\ket{00}$ or $\ket{11}$ state (rather than a generic superposition), then we are restricted primarily to single body potentials.  However, once we allow a solution of the form $\left( \ket{00} + \ket{11} \right) / \sqrt{2}$, which produces equally valid exact solutions for the classical problem, then we may start to consider potentials that have interactions, moreover there is now some finite measure of state space that produces valid solutions rather than single points.  

From a theoretical point of view, this draws an interesting distinction between attempts to probabilistically prepare $\ket{00}$ and $\ket{11}$ exactly in turn, which we have seen to be impossible with only queries to an interacting Hamiltonian.  Hence entanglement plays the role here of allowing the solution of interacting potentials with fewer queries, while sticking to a hypercube graph Laplacian.  For the specific choice of the maximally entangled solution, which we denote here as $\ket{\psi(z^*)} = \left( \ket{00} + \ket{11} \right)/\sqrt{2}$ and the hypercube Laplacian, we may repeat the above analysis for
\begin{align}
   e^{-i \gamma f(z)} \ket{+} &= e^{i \beta X_1}e^{i \beta X_2} \ket{\psi(z^*)}.
\end{align}
In this case, we find that $\beta=\pi/8$ permits solutions, and the corresponding principle potential it solves up to sign and shifts in the identity is given by
\begin{align}
    f(z) = \frac{\pi}{4} Z_1 Z_2
\end{align}
which with scaling by $\gamma$ permits the solution of our problem exactly.  In contrast, there is no scaling which allows the exact solution for an unentangled state.  

It is natural to then ask how this concept extends to greater numbers of qubits.  While a total generalization of these results remains an open problem, we find considerable additional insight by examining specific 3 and 4 qubit instances.  Starting with a state compatible with the ground state of a quantum 3 qubit ferromagnet, a GHZ state of the form $\ket{\psi(z^*)} = \frac{1}{\sqrt{2}} \left( \ket{000} + \ket{111} \right)$.  Performing the same analysis, we find that exact solutions are possible with $\beta=\pi/4$ and the principle potential is given by
\begin{align}
    f(z) = \frac{\pi}{4} (Z_1 Z_2 + Z_2 Z_3 + Z_1 Z_3)
\end{align}
which is the potential for a fully connected two-local Hamiltonian on three spins.  However, we find that if one considers the ferromagnet in 1D (i.e. removes the $Z_1 Z_3$ term), a good solution can no longer be found with a single step and the standard graph Laplacian.  However, in this case we can find a simple modification of the graph Laplacian that allows the solution of this problem.  In particular, if one modifies from the standard Hypercube $\bar L_G=X_1 + X_2 + X_3$ to $\bar L_G = X_1 + X_3$ (remove the $X_2$ term), then $\beta=\pi/4$ again allows an exact solution, however the corresponding potential is
\begin{align}
    f(z) = \frac{\pi}{4}(Z_1 Z_2 + Z_2 Z_3)
\end{align}
which solves the 1D problem exactly.  In this sense, having GHZ-like states as solutions can tend to over correlate the desired potential, and this shows a mechanism by which graph Laplacian modification can expand the type of exact solution possible.

More generally, as the number of qubits increases, we find that results are consistent with the results from the preparation of GHZ states via graph state preparation followed by single qubit Clifford rotations~\cite{hein2006entanglement}.  That is, potentials which correspond to the creation of all-to-all graph states (all-to-all ferromagnets for example) and star graphs  are consistent with preparation of GHZ states.  We found this to be true for examples considered with the appropriate choices for scaling in the graph Laplacian.  These results motivate both the modification of the graph Laplacian as well as support the flexibility gained by considering entangled intermediate states for approximation.

\subsection{Problems of scale} \label{sec:ScaleProblems}
Due to the discreteness of exactly solvable potentials and a single choice of the parameter $\gamma$, we determined potentials of the form
\begin{align}
    f(z) = \sum_i \alpha_i Z_i
\end{align}
with $\alpha \in \mathbb{R}$ are in fact not exactly solvable in either $p=1$ QAOA or simple quantum annealing constructions.  Said another way, $p=1$ low-depth optimization algorithms on the hypercube graph, such as QAOA, cannot solve non-interacting spin problems exactly if their coefficients are not uniform in the right way.  This is analogous to the case of gradient descent having difficulties on problems with poor condition numbers, due to fixing the scale of the step.  If one works out the probability of getting the exact solution as a function of $n$, it is easy to see that for any error, it vanishes exponentially in $n$ for cases that are not special in the distribution of $\alpha_i$.  One simple way of seeing this is to recognize that the initial phasing by $\prod_j \exp[-i \gamma \alpha_j Z_j]$ potentially leads to $j$ distinct states.  By unitarity, the subsequent operations on each qubit $\exp[-i \beta X_j]$ then cannot take distinct states to the same solutions beyond a few chance cases, leading to a slight angle misalignment for all qubits, implying an exponentially decaying probability of success when looking for the exact solution.

In some cases one is not constrained to finding only the exact solution, and good approximations are acceptable.  It is then worthwhile to ask how detrimental this problem is for this case.  Consider a set of randomly distributed $\alpha_i$ with probability measure $d\mu(\boldsymbol \alpha)$.  We will assume that the $\alpha_i$ are independently, identically distributed (i.i.d.) such that $d\mu(\boldsymbol \alpha) = \prod_j d\mu(\alpha_j)$, and hence we will denote $d\mu(\alpha_j)$ by $d\mu(\alpha)$.   For a given distribution, we will be interested in the average performance per spin, which allows us to understand how the solution is scaling in a large size limit.  In particular, due to the uncoupled nature of the spins, we know the expected value of the ground state energy $E_0$ per spin is given by
\begin{align}
    f(z^*) &= \lim_{n \rightarrow \infty} \mathbb{E}_{\boldsymbol \alpha} [E_0 / n] \notag \\
    & = \lim_{n \rightarrow \infty}  \sum_{\vec \alpha} \sum_j - \frac{|\alpha_j|}{n} d \mu(\boldsymbol \alpha) \notag \\
    &= -  \lim_{n \rightarrow \infty} \frac{1}{n} \sum_j [\sum_{\alpha} |\alpha| d \mu(\alpha)] \notag \\
    & = -\sum_{\alpha} |\alpha| d \mu(\alpha).
\end{align}
From symmetry in the Hamiltonian, we know that the maximum energy $A_{\text{max}} = -A_{\text{min}}$.  We also know that for the system we consider, the overlap for a single spin with a given $\beta, \gamma$ is given by
\begin{align}
    O(\gamma, \beta) &= \sum_\alpha 
    |\bra{z^*}e^{-i \beta L_G}e^{-i \gamma f(z)} \ket{+}|^2 d\mu(\alpha) \notag \\
    &= \sum_\alpha  \frac{1}{2} \left[ 1 - 2 \sin (2 \beta) \sin (2 |\alpha| \gamma) \right] d\mu(\alpha)
\end{align}
The expected energy for a single round of QAOA on the other hand is given by
\begin{align}
    & C_m(\gamma, \beta) = \lim_{n \rightarrow \infty} \mathbb{E}_{\boldsymbol \alpha} [\bra{\gamma, \beta} C/n \ket{\gamma, \beta}] \notag \\
    &= \lim_{n \rightarrow \infty} \sum_{\boldsymbol \alpha} \sum_j \frac{\alpha_j}{n} \bra{\gamma, \beta} Z_j \ket{\gamma, \beta} d \mu(\boldsymbol \alpha) \notag \\
    &= \lim_{n \rightarrow \infty} \sum_j \frac{1}{n} [\sum_\alpha \alpha \bra{\gamma, \beta} Z_j \ket{\gamma, \beta} d \mu(\alpha)] \notag \\
    &= \sum_\alpha \alpha \bra{\gamma, \beta} Z \ket{\gamma, \beta} d \mu(\alpha) \notag \\
    &= \sum_\alpha \alpha \sin [2 \beta] \sin [2 \alpha \gamma ] d \mu(\alpha)
\end{align}
For examining approximate solution quality, we will also be interested in the approximation ratio specific to the optimal choice of parameters, defined here as
\begin{align}
 R(\gamma, \beta) = \frac{f_\text{max} - C_m(\gamma, \beta)}{f_\text{max} - f(z^*)}
\end{align}

We now consider probability distributions for $\alpha$ with mean $0$.  Specifically, we consider the discrete binary distribution $d\mu^b(\alpha)$ with $d\mu^b(-1) = d\mu^b(1) = 1/2$, the continuous uniform distribution $d\mu^u(\alpha) = \frac{1}{2} d\alpha$, with $\alpha \in [-1, 1]$, and the Gaussian distribution with variance 1, $d\mu^g(\alpha)=(1 / \sqrt{\pi}) \exp(- \alpha^2 )d\alpha$.  

\begin{figure*}[t!]
\begin{center}
\begin{tabular}{ c|c c c c c c } 
Distribution $d\mu(\alpha)$ & $C_m(\gamma, \beta)$ & $f(z^*)$  & $\gamma^*$ & $C_m(\gamma^*, \beta^*) $ & $O(\gamma^*, \beta^*)$ & $R(\gamma^*, \beta^*)$\\
\hline
Binary & $\sin (2 \beta) \sin (2 \gamma)$ & $-1$ & $-\pi/4$ & $-1$ & $1$ & $1$ \\
Uniform & $\frac{1}{4 \gamma^2} \sin 2 \beta (\sin 2 \gamma - 2 \gamma \cos 2 \gamma)$ & $-1/2$ & $-1.04$ & $-.436$ & $.858$ & $0.936$ \\
Gaussian & $e^{- \gamma^2} \gamma \sin 2 \beta$ & $-1/\sqrt{\pi}$ & $-1/\sqrt{2}$ & $-1/\sqrt{2 e}$ & 0.789 & 0.88\\ 
Annealing Gaussian $(v=1)$ & & & & & & 0.88
\end{tabular}
\end{center}
\caption{Average performance per qubit of a single step of QAOA over distributions of uncoupled random spins.  Each quantity is reported for the intensive value of the average, which eventually concentrates in the large $n$ limit.  The worst performance examined is for a Gaussian distribution of couplings with unit variance.  In this case the probability of finding an exact solution decays exponentially in $n$ as given by $O(\gamma^*, \beta^*)^n \approx 0.8^n$.  Curiously, the approximation ratio $R(\gamma^*, \beta^*)$ achieved in this case is the same as achieved by an annealing protocol with a ramp speed of $v=1$ (Annealing Gaussian $(v=1)$), detailed in the text \label{fig:DistributionPerformance}.}
\end{figure*}

If we integrate over these measures for the above quantities, we find the performances summarized in Fig. \ref{fig:DistributionPerformance}.  While for the binary measure a single round of QAOA is capable of achieving a perfect solution, the uniform distribution is not quite as simple.  In this case, it is clear we may choose again $\beta^*=\pi/4$, however the exact numerical optima is at $\gamma^*=-1.04$ with an optimal value of $C_m(\gamma^*, \beta^*) \approx -0.436$.  A typical method of reporting is the approximation ratio $R^*$, which for the binary distribution is $1$, but for the uniform distribution is $(1/2 + 0.436) = 0.936$.  For the Gaussian distribution, we may also seek a numerical solution, and we find that for $\beta=\pi/4$, the optimum is at $\gamma=-1/\sqrt{2}$, with a value of $C_m(\gamma^*, \beta^*) =-1/\sqrt{2 e} \approx -.4288$.  This yields an approximation ratio of $R^* = (1/\sqrt{\pi} + 1/\sqrt{2 e}) / (2/\sqrt{\pi}) = 1/4(2 + \sqrt{2 \pi / e}) \approx 0.88$. For comparison, it was found that the optimal value for the Sherrington-Kirkpatrick spin-glass model with Gaussian random coefficients had an value of $C_m^{\text{sk}}(\gamma^*, \beta^*)=-1/\sqrt{4e}$, which if one assumes an approximately symmetric spectrum yields an approximation ratio of $R^*= (0.763166 + 1 \sqrt{4 e})/(2 \times 0.763166) \approx 0.70$~\cite{farhi2019quantum}.  

If we turn our attention to the values of $O_{\gamma, \beta}^{g*}$, we see that for Gaussian random coefficients, the probability of getting the exact solution in the large $n$ limit, is likely to scale as $\approx 0.8^n$, which rapidly diminishes as a function of $n$.  While this is better than random guessing for modest values of $n$, it is still troubling for a problem with a trivial solution.  Given that the query to the objective function in a non-interacting problem on a quantum device requires the same complexity as finding the perfect solution on a classical device, it is worthwhile using the intuition behind this failure to motivate methods for improvement.

In classical methods this is often done with approximate inverses of the Hessian or more general preconditioners.  As we are considering a global rather than local optimization, it is more difficult to re-scale aspects of the problem without changing the solution by accident.  While it would be ideal to simply map the graph $G$ to a related one which is independent for each qubit, as this preserves the exact solution, it's easy to see from the exact constraints above, this actually doesn't work.  There is a fundamental asymmetry in the parameters of the potential and the graph, which draws another stark contrast with traditional interpretations of gradient descent.  

The prototypical interacting problem where a vanishing gap is caused by problems of scale was introduced by Fisher~\cite{fisher1995critical} with later work by Reichardt~\cite{reichardt2004quantum}, and interestingly can be mapped exactly to our problem of scale with non-interacting spins for the case of $p=1$ QAOA.  This problem is a simple 1D ferromagnet, defined by
\begin{align}
    H = \sum_i \frac{J_i}{2} (1 - Z_i Z_{i+1})
\end{align}
where $J_i \in \{1, 2\}$ are chosen randomly.  This problem has a gap that scales as $\exp\left(-c \sqrt{n}\right)$ for a small constant $c$ despite both the lack of conflicting terms and the clear path to either of the optimal solutions ($\ket{0}^{\otimes n}$ or $\ket{1}^{\otimes n}$.  These closures happens intuitively because large domains of $J_i$ freeze at different rates, causing conflict between the domains, preventing the solution from settling to the correct value.  

For $p=1$ QAOA, through a fermionic representation of this Hamiltonian in the periodic case it has been shown that the the ring of disagrees maps exactly to a system of uncoupled spins with coefficients determined by momenta in the transformed problems~\cite{wang2018quantum}.  Hence degradation of performance in this model for small values of $p$ is explained by problems of scale in a non-interacting spin picture.  Moreover, this transformation is suggestive of an ansatz for VQE that allows a convexification of the potential surface.

\subsection{Scale in annealing}

Taking a detour from the digital optimization framework, it is interesting to ask how the concerns of scale affect analog quantum annealing approaches.  Motivated partly by hardware constraints, in this setup, we have a continuous time evolution with a time dependent Hamiltonian
\begin{align}
    H_a(t) = f(z) - \Gamma t L_G
\end{align}
where we have assumed a linear ramping of the graph Laplacian sometimes referred to as transverse field due its physical origin in implementation.  In quantum annealing, one imagines starting with the standard $\ket{+}^{\otimes n}$ state, compensates for the always-on problem $f(z)$ through the use of a strong field $\Gamma t$ with $|t| >> 1$.  This makes the problem amenable to approximate analysis through Landau-Zener transition theory.  We imagine starting in the $t \rightarrow -\infty$ limit, ramping at rate $\Gamma > 0$, and counting success as remaining in the adiabatic ground state.  For the case of uncoupled spins, the minimum gap is given by $2 |\alpha_i|$ for an isolated spin.  Hence for a single spin, we may compute a Landau-Zener probability of success as
\begin{align}
    P_{LZ} = 1 - \exp(- \pi \alpha_i^2 / \Gamma)
\end{align}
which for a Gaussian distribution over $\alpha_i$, $d\mu^g(\alpha)=(1 / \sqrt{\pi}) \exp(- \alpha^2 )d\alpha$ dictates an average success probability of
\begin{align}
    O^g_{LZ} &= \int d\mu^g(\alpha) [1 - \exp(- \pi \alpha^2 / \Gamma)] \notag \\
    &= 1 - \sqrt \frac{\Gamma}{\Gamma + \pi}.
\end{align}
In the large $n$ limit, due to independence this implies the average probability of a correct solution scales like $(O^g_{LZ})^n$, which implies to maintain a constant success probability $q$, we have
\begin{align}
    \Gamma &< \frac{(1 - q^{1/n})^2}{(2 - q^{1/n}) q^{1/n}} \notag \\
    & \approx (1 - q^{1/n})^2 \notag \\
    & \approx (1 - e^{(\log q) / n} )^2
\end{align}
which decays exponentially in $n$ for a fixed probability of success, implying that the rate $\Gamma$ must be taken exponentially small in the number of qubits. For this model where the algorithm run time scales as $1/\Gamma$, this implies exponential time resources. As before, sometimes we discard the notion of exact solution, and look for the approximation ratio to guide what fraction of the problem we are getting correct.  For this, the relevant quantity as before is given by
\begin{align}
    A^g_{LZ} &= \int d\mu^g(\alpha) (-|\alpha|) [1 - \exp(- \pi \alpha^2 / \Gamma)] \notag \\
     &=- \frac{\sqrt{\pi}}{\pi + \Gamma}
\end{align}
which yields an approximation ratio of $(2\pi + \Gamma)/(2(\pi + \Gamma))$, which has the correct limit of $1$ as $\Gamma \rightarrow 0$.  At $\Gamma=1$, we have $\approx 0.879$, which is nearly identical to the result of $p=1$ QAOA for the same uncoupled spin problem.

\section{Continuous extensions, relaxations, and rounding} \label{sec:ContinuousExts}
By parameterizing the solution of discrete optimization problems with continuous parameters in quantum circuits, methods like VQE or QAOA are mapping a discrete optimization problem to a continuous, but often non-convex, problem.  Solution of discrete optimization problems by mapping to continuous constructions is not an approach unique to quantum computing, and here we try to draw out some of the connections to classical methods of continuous extensions and relaxations in order to encourage cross pollination of insights and analysis between the two communities.  At least one technique used in the field of classical continuous extensions, known as iterative rounding, will be especially useful in the remedy of problems of scale that were discussed.

The use of continuous extensions is often discussed for classical problems in the context of set functions, which map a set of items $A \subseteq S$ to a value $f(A)$, where $A$ might be the set of edges in a max-cut problem for example.  Such set functions can be mapped to a binary representation where $1$ represents the inclusion of an element in the set $S$, which brings us to a familiar bitwise representation $f(z)$ we've been using this work so far.  A continuous extension of the function $f(z)$, where $z_i \in \{0, 1\}$ is a function $\hat f(x)$ where $x_i \in [0, 1]$ where the function agrees on the vertices of the hypercube (where $x_i$ is exactly equal to $0$ or $1$).

Of particular interest in classical continuous extensions is when a function has a property that roughly corresponds to the addition of elements having diminishing returns (more formally the property of sub-modularity defined by $f(A \cup \{j\}) - f(A) \geq f(B \cup \{j\}) - f(B)$ for $j \in S$ and subset relationship $A \subseteq B \subseteq S$), a simple continuous extension known as the Lov\'asz extension may be formulated that maps sub-modular functions exactly to convex continuous optimizations, which coincides with its convex closure.  This mapping allows efficient, exact solution of some instances of that display sub-modularity.

Importantly for our discussion, the Lov\'asz and related extensions can be formulated exactly as using parameterizations of probability distributions and marginals for the values of the individual bits.  This is exactly what we use the quantum device to do in quantum approaches such as VQE or QAOA, where the quantum wavefunction encodes a probability relationship between the bits and samples that produce candidate solutions.  Consider a trivial VQE circuit for quantum optimization given by
\begin{align}
    \ket{\psi(\theta)} = \prod_j^n e^{-i \theta_j Y_j} \ket{0}^{\otimes n}.
\end{align}
When evaluated on the objective function $\bra{\psi(\theta) }f(z) \ket{\psi(\theta)}$, this expectation value agrees with the discrete form of the problem at all vertices of the hypercube, and is a continuous function that has no solutions lower in energy than the true solution.  Hence it is a continuous extension of the discrete optimization problem.  In fact this particular continuous extension is directly related to the so-called multi-linear extension, defined on set functions $f(A)$ by 
\begin{align}
    \hat f(x) = \sum_{A \subseteq S} f(A) \prod_{i \in A} x_i \prod_{i \notin A} (1 - x_i)
\end{align}
where $x_i \in [0, 1]$ represent a marginal probability of including element $i$. If we take each qubit in the state $\ket{\psi(\theta)}$, parameterize it as $\ket{\psi(\theta)}_i = \sqrt{1 - |c_i(\theta)|^2}\ket{0} + c_i(\theta) \ket{1}$, and make the correspondence $x_i=|c_i|^2$, then we find that the expectation value $\bra{\psi(\theta) }f(z) \ket{\psi(\theta)}$ is exactly the multi-linear extension of the function $f(z)$ in its bitwise representation.  Hence this variational ansatz given is a simple re-parameterization of the multi-linear extension.

From its product construction we know that this ansatz will not provide a quantum advantage, however this connection does let us connect with some results from the classical literature.  In particular, for sub-modular functions, the multi-linear extension gives an extension that is cross-convex.  We do not get into the details of this here, but it has implications for allowing efficient solutions of non-trivial classes of problems~\cite{dughmi2009submodular,vazirani2013approximation}.  As ansatz capable of generating entanglement generally contain the product distribution as a restriction, there is hope that slight perturbations of these problems may also be solvable with more developed ansatz.

The Lov\'asz extension, the multi-linear extension, and the VQE ansatz given here have no dependence on the underlying function.  In the study of extensions, this is termed an ``oblivious'' extension.  In contrast, QAOA and adiabatic approaches fundamentally use the problem $f(z)$ as part of the ansatz, and hence is a non-oblivious mapping of the problem to continuous space.  Moreover, for finite values of $p$, it is not always possible to express the vertices of the hypercube exactly (not all bitstrings are accessible with unit probability), and hence QAOA is not rigorously a continuous extension of the underlying problem nor related by simple parameterization to a continuous extension.  A simple modification of the QAOA algorithm that replaces the preparation of the initial state $\ket{+}^{\otimes n} = H^{\otimes n} \ket{0}^{\otimes n}$ with $\ket{\psi(\theta)}$ allows the approach to function more closely to a continuous extension, and hence adds to the repertoire of possible exact solutions some sub-modular function problems under constraints.

\subsection{Iterated rounding}
When solving the continuous extension or relaxations of a problem, it will sometimes be the case that the best solution found does not correspond to a vertex on the hypercube, but rather some superposition of solutions.  In the classical case, the use of extensions or relaxations are hence often coupled with a ``rounding'' procedure that maps the solution to the nearest vertex~\cite{raghavan1987randomized,vazirani2013approximation,williamson2011design}.  In the field of belief propagation for discrete problems, iterated rounding is termed decimation and more relaxed variations where magnitudes are adjusted are called reinforcement~\cite{mezard2009information}.  In that case, the use of iterated rounding(decimation) can be crucial for proving performance.  While it is fruitful to explore algorithmic variations from all the parallel fields, we will refer to it as iterated rounding here. In fact, for our quantum parameterizations the measurement of bitstrings in the final solution is a randomized variant of this rounding procedure.  Some recent classical approaches to discrete optimization problems work by constructing tensor network approximations to the $\beta-$softened problem of approximating the partition function, $e^{-\beta H}$ taking $\beta \rightarrow \infty$~\cite{rams2018heuristic}.  In this context, a quantum circuit representation of this probability distribution followed by measurement is also a rounding procedure.  While we are fortunate that in the quantum case our continuous relaxations never lead to lower bounds less than the true solution (subject to satisfaction of all constraints), we still require this rounding procedure to give true solutions to our discrete problems.

In the classical case, it has been noted in various contexts that it is often the case that one is close to some of the hypercube vertices much more quickly than others, and it makes sense to round off one of these variables at time, also known as iterated rounding~\cite{lau2011iterative}.  In some cases it is also beneficial for inclusion of hard constraints to round in this way rather than allowing randomized or other deterministic rounding schemes.  In our study of the scaling  of non-interacting problems, the different rate at which each variable approached a vertex was exactly our mechanism of failure leading to potential exponential run times in trivial schedule implementations of quantum annealing and requiring a depth of at least $p>1$ in QAOA for non-interacting problems.  Moreover, if one attempts to use a separable VQE ansatz with a method such as gradient descent, it may also be the case that reaching an exact solution takes more time than desired due to the complexity of reading the gradient out of the quantum device with scales.  Hence it seems that a similar solution may work well in many cases of quantum optimization.

Iterated rounding may be performed in the quantum optimization context as follows.  Consider a quantum optimization circuit $U(\theta)$ that yields the quantum state $\ket{\psi(\theta)}$ from which measurements in the $z$ basis yield potential solutions to the optimization problem $f(z)$.  This $U(\theta)$ could come from a VQE, QAOA, quantum annealing, or other procedure which is adaptive to the problem and number of qubits.  In the collection of these bitstrings, we naturally obtain marginals of $z_i$, $\avg{z_i}$, for the probability of each bit being $0$ or $1$.  In order to avoid pre-maturely locking out optimal solutions on repeated passes at this procedure, we perform the rounding stochastically following an effective inverse temperature value $\beta$, and allow one to set a maximum number of frozen parameters $n_f$.

In each round, we optimize or reconstruct $U(\theta)$ to give an optimal solution, and collect the marginals $\avg{z_i}$.  We then select a variable $i$ to be rounded according to the probability distribution $p(i) \propto \exp[\beta |\avg{z_i} - 1/2|]$. Once the index $i$ is selected for rounding, $z_i$ is then rounded to the closer value with ties broken randomly, and integrated out of the problem by setting this value to be fixed in the optimization problem.  The function $U(\theta)$ is then re-optimized for the new problem, and this process is iterated until upon solution no variables satisfy the constraint set by $n_f$.  In general, both $\beta$ and $n_f$ may be tuned for a problem class of interest as hyperparameters.  We will see in numerical studies that a reasonable choice of these parameters can lead to markedly improved performance for a fixed quantum depth when compared to vanilla rounding by measurement.

This procedure efficiently solves the problems of scale we introduced earlier in this work for all of the procedures examined.  This is because at each step of the procedure, we expect each qubit to well approximate the solution of the qubit of interest despite their products having vanishing support on the exact ground state, and majority voting the qubit results independently would lead to the correct solution.  In more general cases, the process of rounding can represent an approximation controlled by $\beta$ and a number of frozen variables.  We note that a variant of iterated rounding in the context of solving challenges related to $\mathbb{Z}_2$ symmetries in QAOA was recently proposed with $\beta \rightarrow \infty$ and called recursive-QAOA (RQAOA)~\cite{bravyi2019obstacles}.  Given that it solves this additional challenge in the most approximate setting, it gives more support to the notion that iterated rounding can be a powerful improvement to quantum optimization protocols.

\subsection{Identifying opportunity for quantum advantage}
The framework of continuous relaxations gives us an opportunity to speculate as to an origin of possible advantage for variational quantum algorithms for advantage in quantum optimization problems and a potential way to identify this mechanism.  In classical uses of continuous relaxations and rounding, one desires a mapping of the problem to a convex continuous problem that allows a good approximation to the original problem upon rounding.  We saw that the trivial variational ansatz $\prod_j e^{-i \theta_j Y_j}$ is a reasonable choice for some sub-modular problems and trivial uncoupled spin problems.

However, in the quantum case, one usually relaxes the constraint of a convex continuous extension, and instead deals with optimization on the non-convex continuous surface.  Such non-convexity can be addressed in some ways with random restarts or more sophisticated methods for global optimization such as basin hopping.  The use of random restarts must be done carefully to avoid the problem of barren plateaus~\cite{mcclean2018barren,cerezo2020costfunctiondependent} in the optimization landscape however.

For the product ansatz above, it is convex for simple problems with sufficient structure, but in the more general case we expect a rugged and challenging landscape.  If we take the fully parameterized limit, such that a quantum circuit $U(\theta)$ directly and uniquely parameterizes every coefficient, i.e. $\ket{\theta} = \sum_i^{2^n} c(\theta_i) \ket{i}$, as the problem is diagonal, we see that after shift and normalization it amounts to the minimization of a positive-definite quadratic form on a sphere.  Hence, the optimization problem is effectively convex.  

While such a parameterization is obviously inefficient for many reasons, it suggests that between the product ansatz and the fully parameterized limit, the additional directions in the space are breaking down barriers in the landscape making it more convex as additional parameters are added.  If a polynomial number of additional parameters with a polynomial circuit depth widens the basin of attraction of the global solution to a sufficient degree, then the solution of the problem can become efficient with the addition of restarts, even if the space is not completely convex.  As such, for a set of instances or particular instance of a problem, one can study the growth of the basin of attraction with the addition of variational parameters to understand the returns for this type of continuous extension.  We note that the starting points in this space cannot be arbitrary due to the problem of barren plateaus~\cite{mcclean2018barren}, but reasonable seeding schemes can be constructed which well cover the the vertices of the hypercube and other short circuits.

As a simple example of convexification of a landscape through extension, recall the ring of disagrees given by the $1D$ ising model $\sum_i Z_i Z_{i+1}$.  Using a fermionic representation of the problem as done by Wang et. al~\cite{wang2018quantum}, this problem maps exactly to a non-interacting spin problem with differing coefficients.  Using this momentum representation by either transforming the problem classically, or constructing a circuit using this parameterization, this problem becomes a convex optimization on the continuous extension.  In contrast, a QAOA formulation fails in to succeed at $p=1$ in either the original or momentum formulation due to limited ansatz flexibility, and a classical optimization via bit-flips will not yield simple trap-free optimization.  While this problem is not challenging classically, it is suggestive of the way a circuit structure can interplay with a classical optimization problem to facilitate the solution in a variational setting.  More generally, the ability to find a unitary parameterization into an independent spin model suggests the existence of an $n$ parameter extension that renders the optimization convex.

A construction of this form has a close analogy with classical continuous extensions that are formed by neural network architectures~\cite{smith1999neural,bello2016neural,wang2019satnet}.  In this regard, it will be interesting to compare classical neural network extensions with quantum counterparts to understand how the landscape and efficacy are impacted.  It is likely that conclusions in this area will have to be made empirically, but with the rapid development of quantum computers, one that may be testable on real devices soon.

\subsection{Connections to classical homotopy methods}
In drawing connections to classical optimization methods, it is worth pointing out the strong connection to homotopy methods for global optimization (sometimes also called continuation, deformation, smoothing, or embedding methods)~\cite{watson1989modern,addis2006trust,dunlavy2005homotopy,floudas2014recent} or solutions of non-linear equations~\cite{allgower2012numerical}.  These methods work by starting from the solution to an easily solvable problem and deforming that solution slowly to the final problem.  If one deforms slowly enough and certain conditions on the relation between the initial and final problem are met, similar guarantees of solution quality can be found, and with enough time one can even guarantee enumeration of all local optima as well. From just the description it is clear that these methods are related to adiabatic methods for optimization on quantum computers, but so far there has been little cross pollination of ideas between these two areas and the challenges they may both face for practical implementation.

In adiabatic quantum computation the path deformation is performed entirely on a quantum computer, and hence we expect some differences.  However, methods that seem to blend the two together are easy to formulate.  For example, the separable ansatz we have introduced can be left on the quantum computer, and a homotopy deformation of the form introduced for ``adiabatically assisted'' VQE~\cite{garcia2018addressing} can be used to immediately connect the two.  In the case of a separable ansatz, it's clear that the work being done is almost entirely classical, while in the case of adiabatic quantum computing the work is entirely quantum.  However, as one extends ansatz to depths that are difficult to simulate classically, and performs global deformations on the quantum landscape, the interplay becomes more complex.  We believe this is an exciting intersection to explore in future work.

\section{Measure-vote mechanism and mean-field formulations} \label{sec:MeasureVote}
Once one has identified a set of potentials that can be solved exactly, it is natural to look for families of potentials that yield high quality, approximate solutions, and are more closely related to challenging problems of interest.  One family of potentials that are particularly amenable to solutions with low-depth quantum optimization, are potentials that precisely agree with exactly solvable potentials over a non-vanishing fraction of space.  That is, due immediately to linearity, one may imagine the quantum circuit as polling the whole domain of the function at once, and the measure of the domain that agrees the solution is at the optimum determines the quality of solution, independent of how violently the potential disagrees on the other parts of space.   If we take the function $f(z)$ and divide it into an exactly solvable portion $f_{ex}(z)$ given parameters $\gamma^*$, and $\beta^*$ with solution $\ket{z^*}$ that agrees with the solution of $f(z)$ and a perturbation $g(z)$, which is non-zero on only a fraction $c < 1/2 $ of the domain, that is $f(z) = f_{ex}(z) + g(z)$, we can see that
\begin{align}
  & |\bra{z^*} e^{-i \beta^* \bar L_G} e^{-i \gamma^* f(x)} \ket{+}^{\otimes n}|^2 \\
  &= |\bra{z^*} e^{-i \beta^* \bar L_G} e^{-i \gamma^* f_{ex}(z)} e^{-i \gamma g(z)} \ket{+}^{\otimes n}|^2 \notag \\
  &= |\bra{+}^{\otimes n}e^{-i \gamma^* g(z)} \ket{+}^{\otimes n}|^2 \notag \\
  &= |\frac{1}{2^n} \sum_z \bra{z}e^{-i \gamma^* g(z)}\ket{z}|^2 \notag \\
  &= |(1-c) + \frac{1}{2^n} \sum_{z \in D} \bra{z}e^{-i g(z)}\ket{z}|^2 \notag \\
  &\geq |(1 - c) - c|^2 \notag \\
  &= |1 - 2c|^2 \notag 
\end{align}
where we used the definition of $g(z)$ being non-zero on the domain $D$, and $0$ on at most a fraction $c < 1/2$ of the bitstrings to show this immediately implies overlap with the target solution, independent of the values or norms of $g(z)$ on other bit strings.  While the assumption of $c<1/2$ was used to make the analysis simple, tighter results for smaller $c$ have been shown for specific problems, such as Hamming symmetric potentials~\cite{bapat2018bang}.  If the fraction of bitstrings, $c$ is independent of problem size, then solutions of these problems are efficient since we at most need to repeat the procedure a few times to find the exact solutions.  We refer to this particular mechanism of solution as a measure-vote solution, also depicted in Fig. \ref{fig:OverviewCartoon}.  For potentials admitting this class of solution, one often does not need to invoke variational optimization, as one may choose the parameter that solved the unperturbed problem, and by definition the efficiency of solution follows from linearity.  One can also reason that a similar conclusion holds in the more general case that the norm of the perturbation operator $g(z)$ is small from properties of unitary evolution. 

This measure-vote mechanism of solution is exactly what is used in two of the classical problems that are studied in low-depth quantum optimization, the spike problem and bush of implications~\cite{farhi2002quantum}.  These problems are perturbations to the exactly solvable Hamming distance symmetric, or uniform non-interacting spin potentials we identified earlier as being solvable by the Laplacian on the hypercube graph.  They are defined by Hamiltonians
\begin{align}
    H_{\text{spike}} &= w + s(w) \\
    w &= \frac{1}{2} \sum_i (I - Z_i) \\
    s(w) &= \left\{ 
    \begin{array}{c} 
    n^b \text{ if } h \in \frac{n}{4} \pm \frac{n^a}{2} \\
    0 \text{ otherwise }.
    \end{array} \right. \\
    H_{\text{bush}} &= P_0 + (1-P_0)w 
\end{align}
where $w$ is the Hamming weight of a bitstring that we've recalled in the form of Pauli operators, $a$ and $b$ are parameters defining the spike width and height, and $P_0$ is the projector onto the first qubit being $0$. These problems are of great interest, as in certain regimes they are provably hard for the more demanding quantum adiabatic optimization and simulated annealing algorithms, while admitting constant depth solutions when examined using the measure-voting mechanism regime.  Specifically, the complexity separation between measure-voting, simulated annealing, and the quantum adiabatic algorithm was studied in depth by Bapat and Jordan in the context of bang-bang control, and shown to admit exponential time advantages over these competing methods~\cite{bapat2018bang}.  The precise conditions when these problems are hard are enumerated by Bapat and Jordan, but at a coarse level, the spike problem is hard when the spike is tall and wide enough $(2a + b > 1)$, and the bush problem is hard without a modification of the graph Laplacian.  Quantum adiabatic optimization suffers from an gap closure in the case of the spike problem, while the measure-vote mechanism mostly ignores the spike as it occupies a small fraction of the overall domain.

These results look quite promising for low-depth optimization, suggesting that the measure-vote mechanism may be an excellent reference for understanding how quantum interference can provide dramatic advantages for classical optimization problems.  However, we saw that the hypercube Laplacian was most strongly associated with exact solutions on uniform uncoupled spin problems, as evidenced again here by the reference problem being given by $h$, which suggests that for single steps the mechanisms may not be hard to simulate classically, similar to a conjecture made by Bapat and Jordan.  

To examine how dependent the measure-vote mechanism is on underlying quantum effects, we construct a mean-field variant at the level of qubits of an algorithm that proceeds via application of the cost function and exploration via a fast-forwardable graph to examine how such a low-depth optimization to examine how the algorithm proceeds in the absence of entanglement between qubits.  We note that other mean-field variants of quantum adiabatic optimization, for example simulated quantum annealing, have been explored and found gap closures similar to their entangled counterparts for the spike and bush problems.  Here we are examining more precisely the role of quantum effects in the measure-vote mechanism.

To allow for additional flexibility, we define our mean-field model in a way that doesn't depend on the objective function being expressed simply as a tensor product of bits, but that will reduce to common mean-field methods when this is allowed.  At each timestep, we use the action of a mean-field potential operator followed by the action of a mean-field graph Laplacian, each of which are state dependent.  The mean-field potential and graph Laplacian operators for a qubit $i$, on a state $\ket{\psi}$ are given by
\begin{align}
f^i(z; \ket{\psi}) &=  \text{Tr}_{j \neq i} [ f \ket{\psi} \bra{\psi}] \otimes I_{j\neq i} \\
L_G^i(z; \ket{\psi}) &= \text{Tr}_{j \neq i} [ L_G \ket{\psi} \bra{\psi}] \otimes I_{j\neq i}
\end{align}
where $\text{Tr}_{j\neq i}$ indicates the partial trace over all qubits $j$ not equal to $i$ and $I_{j \neq i }$ is the identity operator on the same qubits, designed to lift the action of these operators formally into the full qubit space.  However, as with many mean-field methods, one may choose to work with the reduced one-qubit objects if desired and the operators are amenable to efficient reduction.  The action of the 1-qubit mean-field operators is used to evolve the state by
\begin{align}
    \prod_k \exp \left[-i \beta \bar L_G^k(z; \ket{\psi}) \right] \prod_j \exp \left[-i \gamma f^j(z; \ket{\psi})\right] \ket{\psi}
\end{align}
which is guaranteed to map product states to product states in the spirit of a mean-field method, preserving the lack of entanglement.  We note that for efficiency, one is able to take advantage of the Hadamard diagonalizability of the graph Laplacians here if desired, as well as the convenient definitions in terms of tensor products of Pauli operators.  While one could lump together a number of qubits into qudits to create variants of mean-field calculations with some entanglement at the level of qubits, we do not explore this here.  For qubit separable Hamiltonians on the hypercube graph, such as the Hamming distance symmetric potential $w$, this mean-field construction is exact.

\begin{figure}[t!]
\centering
\includegraphics[width=8cm]{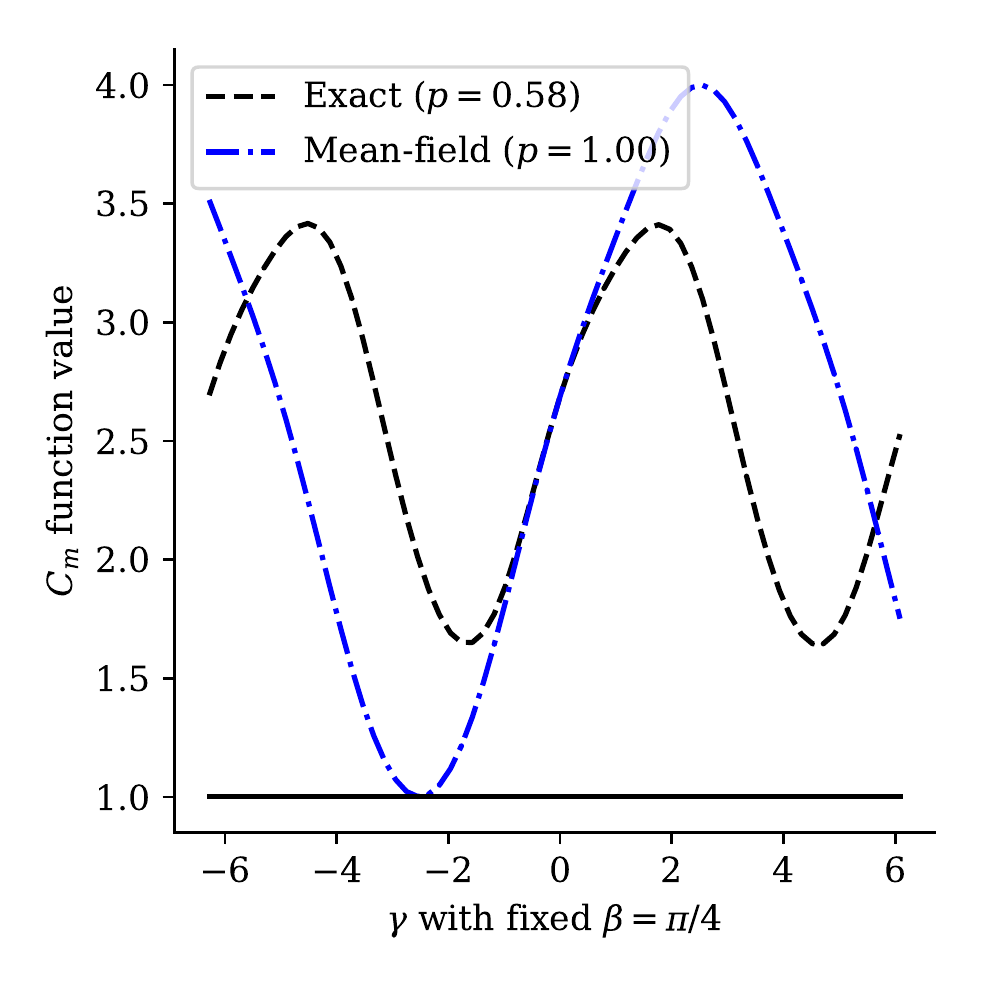}
    \caption{Comparison of mean-field and exact single step for spike potential.  As the prototypical example for the success of low-depth optimization strategies over quantum adiabatic optimization, this represents an important case that exemplifies the measure-vote mechanism. The number $p$ behind each method in the label represents the overlap with the exact solution found under a non-local search of the parameters $\gamma, \beta$ (off-chart). The distinction between the mean-field and exact case implies entanglement is involved in the mechanism, but is strictly detrimental to performance, due to the exact solution coinciding with the solution of the uncoupled spin problem at the foundation of the spike.
    \label{fig:MeanField}}
\end{figure}

We apply this construction to both the spike potential and the bush of implications, and summarize the results in Fig. \ref{fig:MeanField}.  For a measure-vote mechanism, as we emphasized in our discussion of exact results, the $\beta=\pi/4$ cross-section is most important, so we plot the objective function (expected value of the cost function here) for different values of $\gamma$ along this line for both the exact and mean-field constructions.  The dramatically different shape of the landscapes implies that quantum correlations are playing a role.  However, it is impossible not to notice that the performance of the mean-field method markedly exceeds that of the exact case with respect to solving the actual problem.  Hence for these exact instances of study, while entanglement is involved, it appears to be strictly detrimental for the purposes of solving the optimization problem.

\begin{figure}[t!]
\centering
\includegraphics[width=8cm]{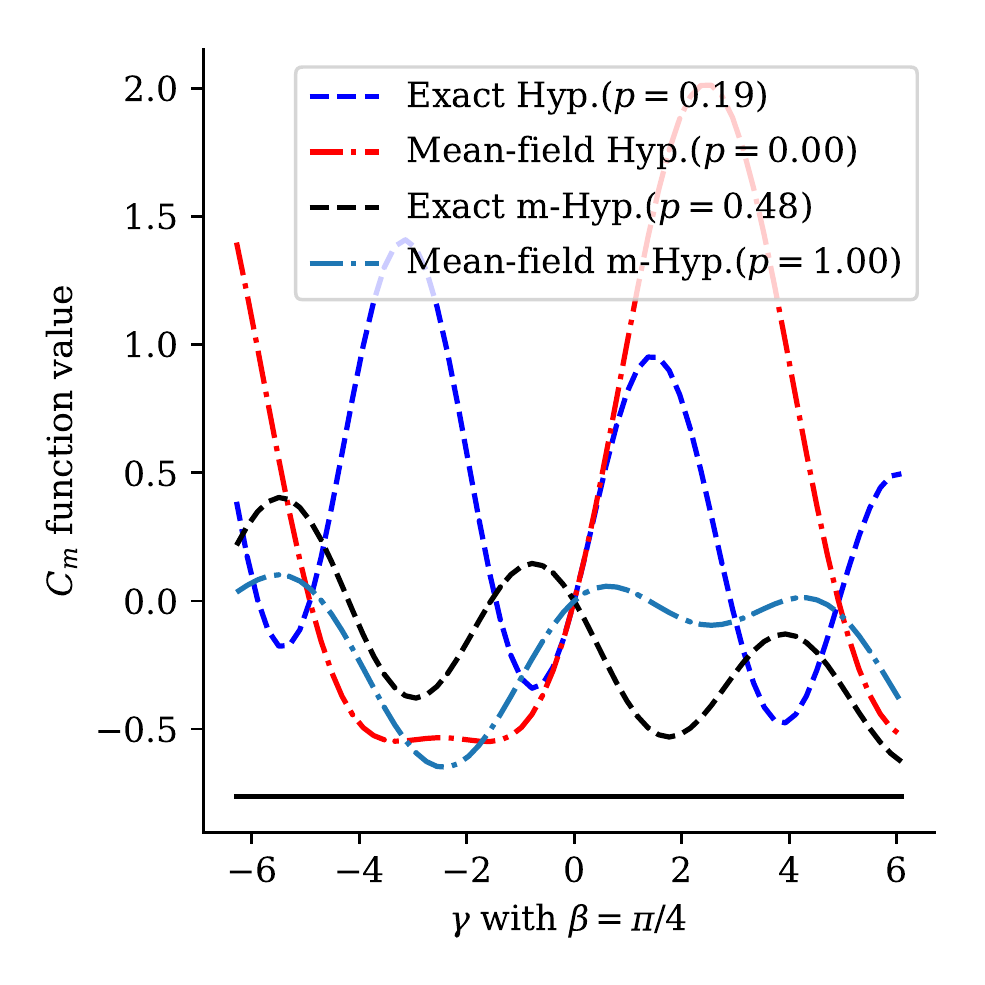}
    \caption{Comparison of mean-field and exact single step for the conflicted pair potential.  Hyp. and m-Hyp. refer to the hypercube graph Laplacian and a modified Hypercube where the corresponding coefficient is shifted to $1$ to $3$.  The number $p$ behind each method in the label represents the overlap with the exact solution found under a non-local search of the parameters $\gamma, \beta$ (off-chart).  We see that the flexibility offered by individual modification of graph Laplacian terms allows dramatically improved solutions in this case, although entanglement is not seen to be overwhelmingly beneficial.
    \label{fig:PairMeanField}}
\end{figure}

To understand why the mean-field case exceeds the performance of the exact in this case, it is helpful to identify a peculiarity of a single step of this algorithm starting from the $\ket{+}^{\otimes n}$ state.  In particular, when one evaluates the contribution of a multi-qubit $Z$ operator, e.g. $Z_1 Z_2$, to the mean-field potential in the first step, we find that
\begin{align}
    \text{Tr}_{j \neq i} [ Z_i Z_j \ket{+}^{\otimes n} \bra{+}^{\otimes n}] \otimes I_{j\neq i} = 0  
\end{align}
for all $j$, and this result holds more generally for the case where one has equal population on the $\ket{0}$ and $\ket{1}$ state of the $k$'th qubit, independent of their relative phases.  Hence, in the first step, the contribution of all higher body operators is $0$, and the potential is exclusively constructed from the single body operators, $w$, which point to the exact solution by construction, without the confounding impact of the higher body perturbations.  Hence, this mean-field construction is even more robust to non-exact perturbations, but if they influence at all the desired solution, it does not capture this effect.

This result alone does not rule out the possibility of a measure-voting mechanism being productive in quantum optimization settings due to the robustness of the mechanism, however it implies we may need to look for graph Laplacians that are not as strongly associated with separable problems.  It also suggests that there may always be room in optimization portfolios for quantum-inspired algorithms like this mean-field variant, as entanglement may not be universally helpful for all problem instances, even when they are contrived to fix the quantum mechanism of interest as they were here.

\subsection{Conflicted pair potentials} 
While the insights from objectives that are mostly separable provide great insight into the mechanisms of these optimization methods, one is also interested in what happens when the objectives are quite different from this.  To this end, we construct a strongly coupled, conflicted pair potential.  We would like to keep the potential simple enough to reason about exactly, while not finding that vanilla mean-field had superior performance.

To this end, we introduce the conflicted pairs potential
\begin{align}
    f(z) = \sum_i -(1 + \epsilon) Z_{2i} - Z_{2i + 1} + \delta Z_{2i} Z_{2i+1}
\end{align}
where $\delta > 2 + \epsilon$ to allow the two-body term to dominate and $\epsilon >0$ takes a small value to break the degeneracy of the ground state but remain close to a one-body potential that would be exactly solvable.  This is a conflicted potential in the sense that the ground state of the two one body terms disagrees with the ground state of the two-body term, which ultimately determines the ground state of the full Hamiltonian.  As the two-body term dominates, the mean-field methods of the last section fail completely without modification of the Laplacian, and this is a good test ground to go beyond the basic construction from before.  As this problem is a simple sum of uncoupled two-body potentials, it also remains easily exactly solvable.

We plot the results from examining this potential in Fig. \ref{fig:PairMeanField}.  A striking difference from the results on the spike potential, is that the mean-field hypercube driver is now significantly worse than other methods.  As we saw in the last section, the mean-field hypercube potential can only read the one-body part of the potential in the first step, and as a result, it cannot find contributions in the phase that lead to the correct state.  In this case, the exact case with entanglement fairs markedly better, though not ideally.  The use of entanglement here allows contribution by the two-body terms, though the one-body terms remain dominant and hamper overall performance significantly.

To understand the power of modifying the graph Laplacian, we examine a modified Laplacian where select qubits shift their coefficient from $1$ to $3$ allowing some terms in the scaled Laplacian to have coefficient $\pi/4$ while others have coefficient $3 \pi / 4$. By examining a mean-field variant of this Laplacian as well, we note that a majority of the conflicted terms can be overcome at the one-body level through a custom, qubit-independent Laplacian. We note that this bears an interesting connection to an earlier work by Farhi et. al, where it was found that stochastically switching some of the graph Laplacian terms between $\pi/4$ and $3 \pi/4$ can open the gap from an exponential worst case in some instances ~\cite{farhi2009quantum}.  We note that in this case, the classical optimization landscape is considerably more complex than previous cases.

\section{Kinetic energy cannot be ignored} \label{sec:KineticEnergy}
One of the earliest lessons taught in quantum mechanics is learned from the particle in a box: confinement of a quantum wavefunction raises the energy of the system.  In connection with this idea, a key distinction between an iterative classical optimization and one proceeding via quantum dynamics, is that decreasing the number of bitstrings that have finite amplitude in the quantum case fundamentally increases the kinetic energy through confinement.  This generates a force that directly opposes our goal of finding a solution that may represent only a single bitstring.  Moreover, phase randomization from prioritizing optimization of the potential can raise the kinetic energy in a way that gaining information from the potential can become problematic.  Here we explore some of the implications and nuances resulting from kinetic energy that lead towards suggested improvements to existing algorithms.

\begin{figure}[t!]
\centering
\includegraphics[width=8cm]{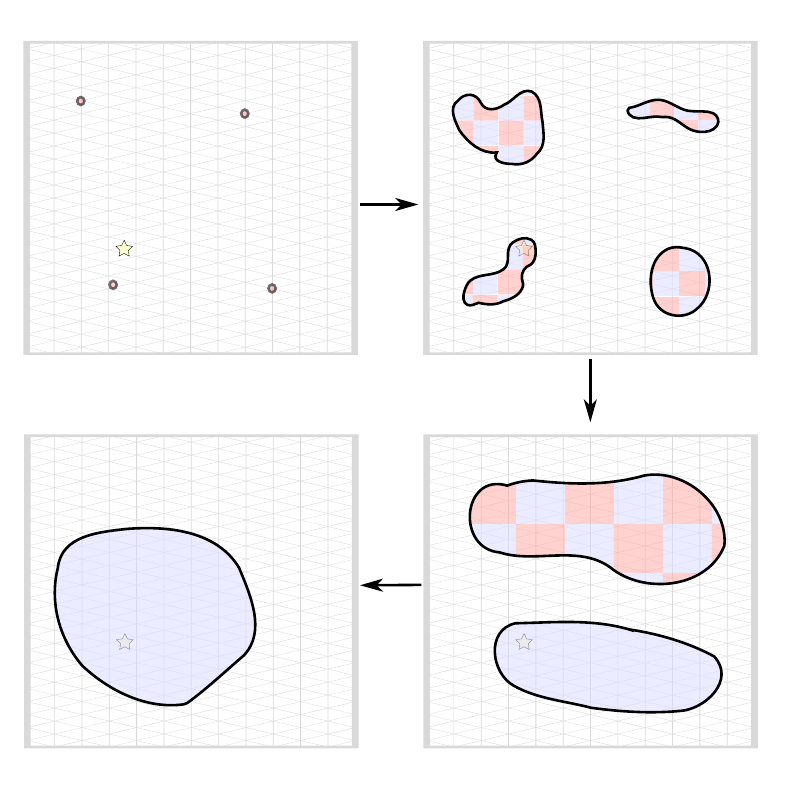}
    \caption{Cartoon of extremes of improvement or information gain.  The arrows follow a depiction of increasing information gain per unit step.  The least information gain from the potential comes from any single bitstring of support or vastly separated single bitstrings.  As one grows the domain of support, or number of disjoint domains of support with randomized phases, the information gain increases.  The existence of large coherent domains (solid colors) allows for that fraction of the space to effectively gain information, while checkered parts often scatter aimlessly.  Finally as phases within domains align and the domains of support become larger, the information gain per step is maximized.  This caricature emphasizes the need for intermediate steps of optimization algorithms to consider phase alignment.
    \label{fig:InformationGain}}
\end{figure}

To get a sense of the kinetic energy for different wavefunctions, one can consider a simple 1D graph of 2 nodes, which has 
\begin{align}
    L_G = \left( \begin{array}{cc} 1 & -1 \\ -1 & 1 \end{array} \right) = I - X
\end{align}
As before, the kinetic energy is minimal at $\avg{T} \propto 0$ when one has an equal superposition, $\ket{+}$, maximal with $\avg{T} \propto 2$ when the nodes have opposite phase, $\ket{-}$, and in the middle when one has some restriction to either of the individual basis state $\ket{0}$ or $\ket{1}$ with $\avg{T} \propto 1$.  This implies that for equal phase superpositions, the larger the number of cuts of edges on the graph required to isolate the wavefunction in bitstring space, or surface area on the graph, the higher the kinetic energy simply for the reason one has cut the graph.  Moreover, for that number of cuts, alternation of the phase maximizes the kinetic energy.  While this is a trivial example, it imparts the correct intuition for understanding the impact of kinetic energy in the efficiency of quantum optimization algorithms as we will see empirically for more interesting examples later.

In the context of our short depth optimization, the effective kinetic energy is given by $T=\beta L_G$.  We now consider the effect of these low-depth optimization primitives on states that are not the kinetic energy ground state.  This could be either an informed initial state, designed to impart prior information to improve the rate of convergence, or progress from an intermediate step of a quantum optimization algorithm.  

The most extreme example of this is the case where one starts from a fixed, but incorrect bitstring, $\ket{b}$.  Perhaps one believes it to be close to the true result, or it is the measurement result of an intermediate step.  In this case, when applying the operator
\begin{align}
    e^{-i \beta \bar L_G} e^{-i \gamma f(z) } \ket{b}
\end{align}
we find that one merely imparts a global phase from the potential, meaning no information from the problem itself enters the problem.  All dynamics and information comes exclusively from the graph Laplacian diffusion, or kinetic energy.  This suggests two extremes.  The first, or more familiar case, where the initial state is given by $\ket{+}^{\otimes n}$ achieves the most information directly from the potential possible in a single step, since there is no baseline contribution from random diffusion or momentum in previous steps.  The second, is that for any select bitstring, no information comes from the potential, and all dynamics are random transport dictated by the graph.  The existence of these extremes suggests that there is a spectrum of efficiencies one can achieve for different states relative to the graphs.  We cartoon this spectrum of different types of states for clarity in Fig. \ref{fig:InformationGain}.

While confinement of a wavefunction may be a substantial contribution, one can also consider the action on non-confined but high energy states.  For example on the hypercube graph, the state $\ket{-}^{\otimes n}$ is the maximum kinetic energy state.  However if one constructs a similar constraint equation for this as an initial state, 
\begin{align}
    e^{-i \gamma f(z)} \ket{-}^{\otimes n} = e^{i \beta \bar L_G} \ket{z^*}
\end{align}
we find in contrast to the confined state that there appear to be solvable potentials.  Although starting from a non-zero kinetic energy state contrasts with a picture designed with adiabatic dynamics in mind, it is a reminder that we are often seeking explicitly diabatic paths for quantum resource efficiency, and that minimum kinetic energy of the state may not be the best direct indicator of solvability.  However minimizing kinetic energy may be a sufficient condition for removal of random momentum, and the highest kinetic energy states are those with random phases rather than the most confined, which reside in the middle of the spectrum.

To be more concrete, consider again the hypercube graph and imagine a situation where we have a contiguous set of vertices $S$ that are all neighbors with equal magnitude quantum amplitudes inside the set, and 0 amplitudes outside the set.  We can divide contributions to the kinetic energy into two parts, relative phase contributions from the interior of the set, and kinetic energy along the surface of the set, represented by bonds from the occupied vertices to the unoccupied vertices.  

We first focus in on the contributions from the perimeter of the set, as this must always be present in any converging optimization algorithm, even if one cleverly initializes the state to have equal phases on the interior.  The problem of minimizing this boundary is the generalization of isoperimetric problems in euclidean space to graphs, which in the general case is quite difficult, but on the hypercube there are a collection of known results.  

The results differ based on whether one considers the minimum number of vertices one must remove to separate the sets, known as the vertex boundary, or whether one counts the number of edges to remove to separate the two sets, or the edge boundary.  The minimum perimeter shape for a fixed set with respect to vertex boundary is known to be hamming balls or very near to hamming balls if the number of vertices in the set cannot satisfy that exactly.  We will be interested in the interior vertex boundary of the set $S$, denoted $\partial_i(S)$, which we define to be the set of vertices in $S$ that are connected to the complement of $S$.   For the hypercube with a Hamming ball of radius $r$, the size of this interior boundary is given by
\begin{align}
    |\partial_i(S)| &= \left( \begin{array}{c} n \\ r \end{array} \right)
\end{align}
These boundaries are convenient from the point of view of analysis, and will be particularly relevant to strategies we use to coherently increase efficiency later.

To get a sense for the impact on efficiency, imagine one has progressed in optimization until the current set of vertices is $2^{n/2}$, or that random sampling of the state could now produce a quadratic advantage over blind random sampling.  In the worst case, these vertices are completely non-contiguous which is possible even as early as a state with support on $2^{n-1}$ states as the hypercube graph is a bi-partite graph.  Any application of $L_G$ can now only grow the set of occupied states reaching at worst a return to the fully occupied state, and the potential can at best shift the support.  In the less extreme case where the bit strings are confined to a hamming ball of radius $n/2$, the contiguous interior can obtain information from the potential, but the exterior is doomed to suffer random diffusion in at least one direction.  This we term boundary scattering and depict in Fig. \ref{fig:OverviewCartoon}.

So far we have discussed only kinetic energy resulting from confinement on the graph.  The highest energy contributions actually come from relative phase differences on the interior of a set of points, and in principle can be the most damaging to information gain from the potential and progress of the optimization.  The interior phases are in some ways analogous to accumulated momentum from the potential in a momentum based optimization, however after many steps of an algorithm prioritizing the potential contributions, one may end up with randomized phases on the interior overwhelming actual contributions, making optimization progress nearly impossible.  This type of contribution we call phase scattering, which is also shown in Fig. \ref{fig:OverviewCartoon}.

To get a better quantification for both of these effects, we construct the following model problem.  We use the objective that is exactly solvable with the hypercube Laplacian, which is a Hamming distance symmetric $f(z) = w = 1/2 \sum_i (I - Z_i)$, and imagine that a quantum optimization algorithm, such as an adiabatic optimization, has stopped at one of several intermediate points.  We consider the case where the set of support has shrunk to a hamming ball of size $r$ around the solution, with uniform and randomized phases.  In principle, this is a strict improvement towards the global solution.  

We quantify the amount of improvement possible from this starting point measured by overlap with the exact solution in the subsequent step compared with the previous step.  To do this, we take the overlap with the exact solution in the initial state $|c_0|^2$, and compare it with the overlap for the subsequent step, $|c_f|^2$, with $\beta$ and $\gamma$ used to minimize the variational objective, set to be the expected value of the objective with respect to the state here.  As a technical note, we utilize a global sampling on a grid of $\beta$ and $\gamma$ values followed by a local optimization to find the optimal values for each.  This is done to address the ruggedness of the parameter landscape for states with varying phases.  We use their difference normalized by the value for an exact solution with the uniform starting state $(1 - 2^{-n})$.  This quantity, 
\begin{align}
  I_{\ket{\psi}} = \frac{|c_f|^2 - |c_0|^2}{1 - 2^{-n}}
\end{align}
the improvement possible for a state $\ket{\psi}$ we use as a proxy for information gain from the potential and efficiency of the algorithm.  This quantity quantifies the maximum improvement in probability of obtaining the correct solution for adding one round of the form $\exp(-i \beta \bar L_G) \exp(-i \gamma f(z))$.  We note this depends on the state, the Laplacian, the potential, and also the variational objective function used in the optimization of $\beta$ and $\gamma$.  Moreover, due to the use of the mean here, negative values of gain will be possible.  We defer discussion of negative values of this gain to the section on shadow defects.

\begin{figure}[t!]
\centering
\includegraphics[width=8cm]{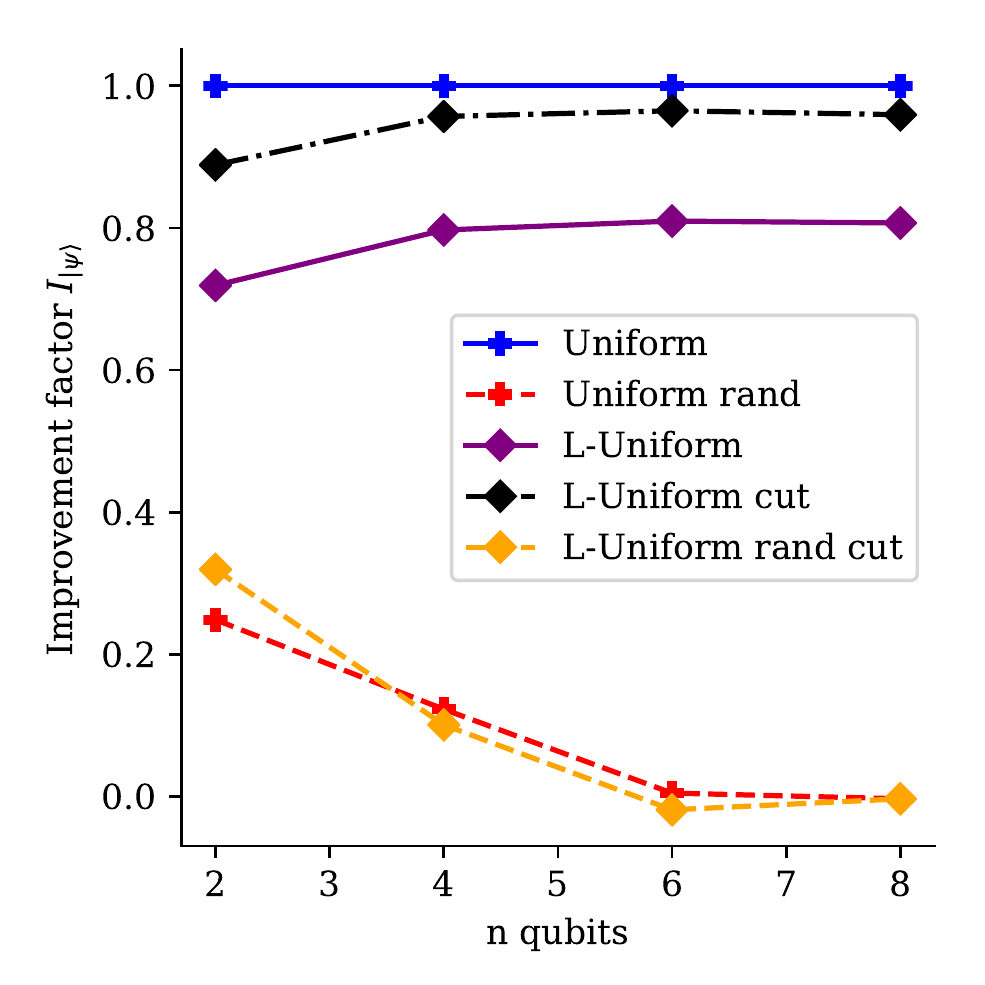}
    \caption{Empirically measured improvement / information gain on different states.  Uniform refers to the standard $\ket{+}^{\otimes n}$, while L-Uniform refers to a localized Hamming ball of radius $n/2$.  The label ``rand'' refers to randomized relative phases on the support region, and cut refers to coherent cutting applied to the graph Laplacian.  Results agree with the developed intuition, that phase scattering destroys the ability of a single step to meaningfully improve.  Boundary scattering plays a smaller but significant role, and much of the effect of boundary scattering is improved by our coherent cutting procedure.
    \label{fig:ImprovementProxy}}
\end{figure}

In Fig. \ref{fig:ImprovementProxy}, we show the results for this proxy measure on the hamming distance symmetric problem for a number of different starting states.  The results are stark, showing that the possible improvement vanishes rapidly for non-uniform states.  Moreover, internal phase randomization seems to render subsequent improvements nearly impossible at larger sizes, suggesting that maintaining phase coherence through the quantum evolution is crucial for success.

There is an interesting connection here to recent work on developing good initial guesses for the QAOA algorithm at higher levels of repetition $p$~\cite{zhou2018quantum,farhi2019quantum}.  It is often found that smooth increase of the potential term and smooth decrease of the kinetic term, akin to an accelerated adiabatic path are often close to the optimal solutions.  While it has been shown in hard cases the mechanism of solution for QAOA may be diabatic~\cite{zhou2018quantum}, our results support the idea that pre-training with kinetic energy and adiabatic solutions in mind facilitates information transfer from the final point to initial stages where the diabatic transition is leveraged.  Moreover, starting from optimizations on fewer layers, and simply adding more layers to the end (rather than interpolating) seems to lead to poor quality local optima.  Our results suggest a direct mechanism for this effect:  greedily optimizing some number of layers for the potential exclusively (as opposed to potential and kinetic) results in phase randomization, which is difficult to improve when taken as a starting point for a subsequent local optimization.

To further this connection, consider the adiabatic path for the hamming symmetric problem.  At each point along the path, the system is in the simultaneous eigenstate of
\begin{align}
    H(s) = -(1-s) X_i - s Z_i
\end{align}
for each qubit, for some value of $s$.  For every value of $s$, the instantaneous eigenstate of this problem is a phaseless superposition of shrinking support.  That is, the adiabatic algorithm follows a relative phaseless Hamming ball of shrinking radius.

In fact, more generally, the structure of using any graph Laplacian $L_G$ as the term in such an adiabatic evolution renders the Hamiltonian stoquastic up to a trivial diagonal shift of the potential.  This implies the instantaneous ground state of the fully adiabatic solution for this type of construction is always phaseless.  Hence, including kinetic energy terms in a variational objectives can provide a way not only to increase general information gain in subsequent steps, but can provide a more reliable witness to diabatic transitions that shortcut one into the instantaneous ground state desired.

Here we have seen that the impact of kinetic energy on the ability to progress in an optimization can be overwhelming.  The correlation of this with results from choosing initial QAOA parameters and the corresponding adiabatic path suggests that a method attempting to improve optimization results must also consider the average of the kinetic energy.  While at any given step, the best result for the optimization is to focus on the objective, this damages the potential for information gain in future steps. This suggests that if one is to use more flexible variational ansatz for optimization problems, the variational objective to train intermediate layers should also include a function of the kinetic energy as well as the potential. 

\section{Shadow defects} \label{sec:ShadowDefects}
The intuitive picture that these short-depth optimizations proceed by measure-vote or some inherently global action over the whole landscape suggests an interesting mode of failure.  It implies that energetic anomalies far from the solution of interest can greatly influence the ability to solve the problem effectively when variational feedback is included.  For example, if for every step via Laplacian diffusion that improved approximation in the global basin, there was wavefunction density near a distant energetic spike, one may conclude falsely conclude that variational improvement is impossible, due to this high energy defect in a remote part of the energy landscape causing an average energy increase.  Here we show some of the simplest examples where this mode of failure occurs, and suggest a partial fix to the problem.  The form of the fix supports recent efforts in changing the cost function for variational quantum optimizations.  It is an interesting direction for future study to understand the relationship between these issues and recently discussed reachability deficits~\cite{akshay2020reachability}. 

\begin{figure}[t!]
\centering
\includegraphics[width=8cm]{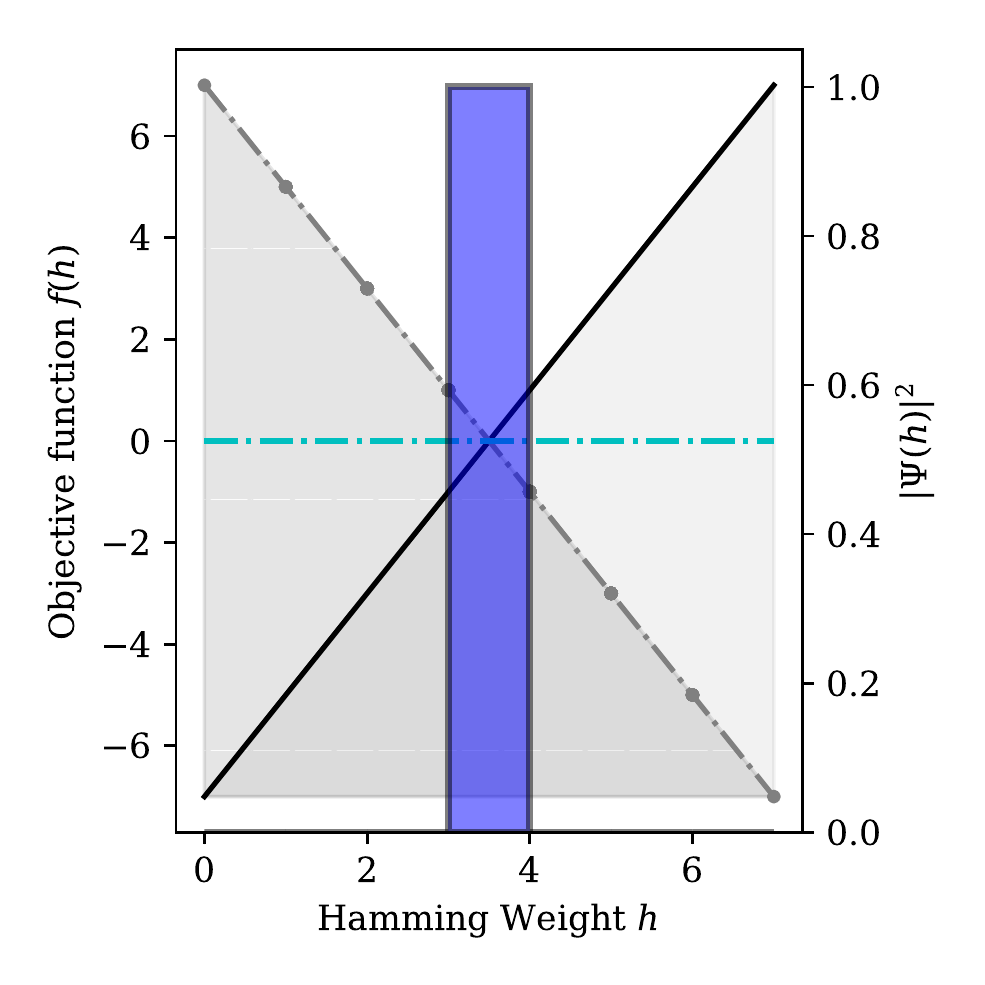}
    \caption{Cartoon of the shadow defect from an exactly solvable potential.  We examine the initial and final state of a low depth optimization step started from the equal superposition of all bitstrings with Hamming distance $n/2$, shown as a blue histogram here.  Even on the exactly solvable Hamming distance ramp, shown in light grey, the variational objective shown as a dotted line in the center is completely flat.  Even though there is no barrier between the initial state and exact solution, this shadow defect cast over the exact solution makes improvement impossible without additional method refinements.  We depict this shadow defect as an artist rendering in grey as the ramp potential in grey on the left of the figure.
    \label{fig:ShadowDefect}}
\end{figure}

Consider again the Hamming distance symmetric problem, $f(z)= w$, with the corresponding hypercube Laplacian.  We imagine that due to some intermediate optimization steps, we have successfully reduced the support of the wavefunction to be closer in Hamming distance to the solution, but not perfectly so.  The wavefunction is equally superposed over all bitstrings with Hamming weight $\lfloor n/2 \rfloor $, for an odd number of qubits.  If one examines the parameter landscape for optimizing over $\beta$ and $\gamma$, using the expected value of the cost function as the optimization metric, the landscape is flat everywhere.  The optimal move is to not move.  Due to the coupled movement of the superposition, every movement downhill is counterbalanced in energy by a movement upwards, as if the uphill part of the potential is casting a shadow defect or barrier on the low energy part.  While this specific case could be related to a symmetry, we term the more general effect as shadow defects, and depict it as well as the initial and final wavefunction, and potential landscape for $\gamma$ in Fig. \ref{fig:ShadowDefect}.  

\begin{figure}[t!]
\centering
\includegraphics[width=8cm]{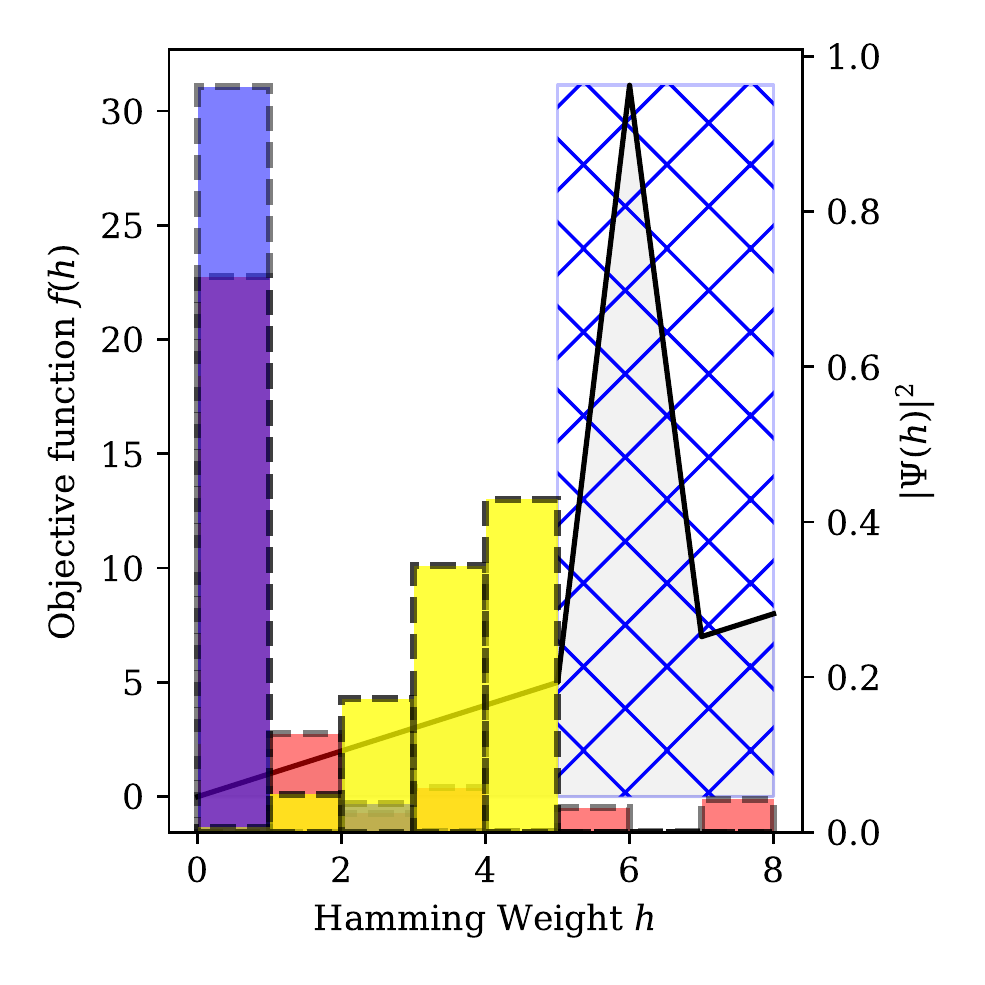}
    \caption{Avoiding spike shadow defects with coherent cuts. In this case, we initialize the wavefunction (yellow) with increased weight on the target state, confined to a Hamming distance of 5 for an 8 qubit system.  The final wavefunction without laplacian cutting (red) degrades in quality at the solution from 0.81 to 0.71 as compared to a case without the spike (grey indicates potential), despite having no support on the spike and a barrier-less path to the solution.  When the graph Laplacian is used to cut (cut indicated in crossed blue, wavefunction in blue) the solution consistently has overlap 0.96.
    \label{fig:ShadowSpikeDefectWithCut}}
\end{figure}

As an additional example, we consider a more realistic case of optimization progress which has narrowed the support down to a Hamming ball around the final solution in Fig ~\ref{fig:ShadowSpikeDefectWithCut}.  In this case, the initial wavefunction with increased support on the final solution (pictured in Yellow) has no overlap with the spike in the potential and in a local method would only see the non-interacting potential.  Yet the final wavefunction in without cutting (red) shows significant degradation in solution quality from $p=0.81$ to $p=0.71$ as compared to the same initial state on the non-interacting potential, with the additional loss coming from boundary effects of kinetic energy.  In contrast, when cutting is applied (removing from the graph laplacian the region in crossed blue), the solution maintains a quality of 0.96 in both the non-interacting and spike case.  Hence these shadow defects, named as they cast a shadow on the landscape rather than obstruct it directly, cause significant degradation of performance due to the global nature of these methods.

\begin{figure}[t!]
\centering
\includegraphics[width=8cm]{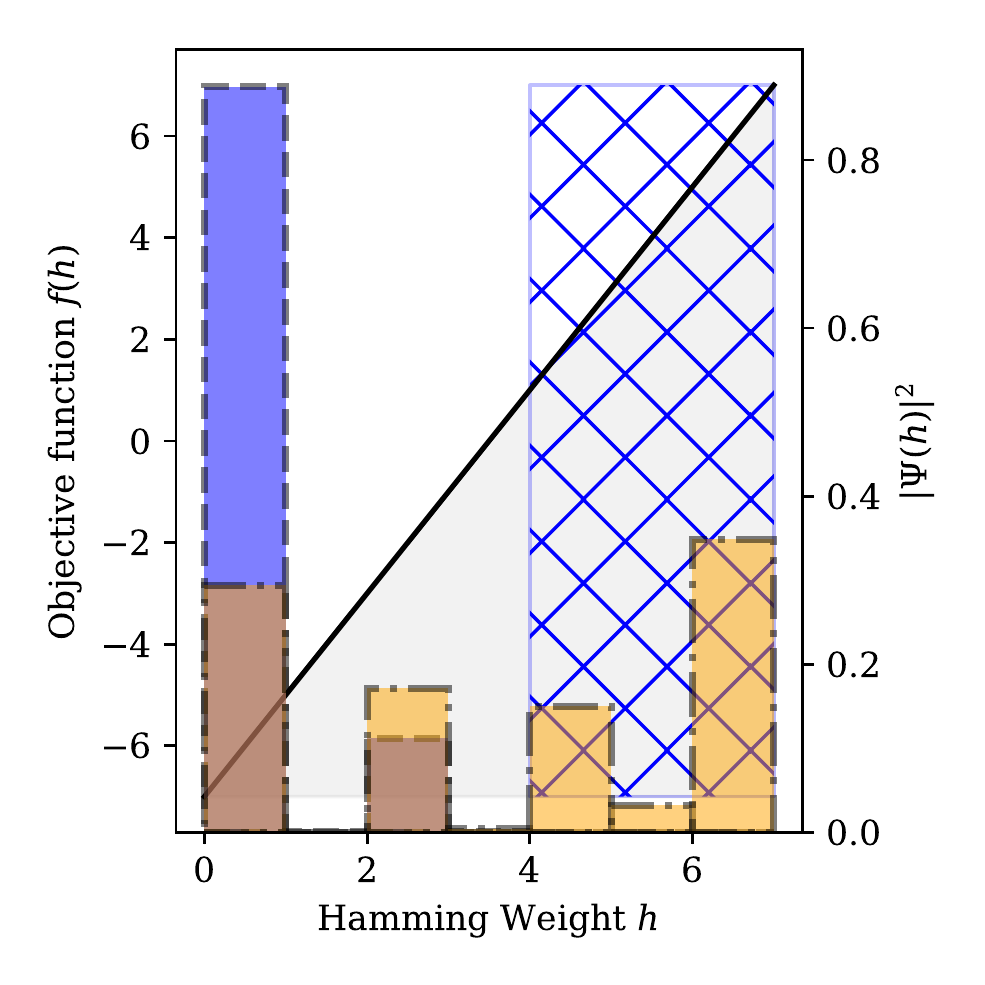}
    \caption{Avoiding shadow defects with coherent cuts and variational objective modification.  Here we show the results for the same initial state and potential that exhibited stalling due to shadow defects.  In blue, we show the final state when using a coherent cutting, where the domain of cutting is depicted in the blue cross hatched box.  The orange result shows the uncut result when switching to the Gibbs variational objective.  Notably this alleviates some of the problem, but suffers from a similar problem of symmetry in ultimate efficiency, suggesting the more scalable solution to shadow defects is cutting the graph Laplacian.
    \label{fig:ShadowDefectWithCut}}
\end{figure}

The ability of distant and unrelated shadow defects to impact the efficacy of low depth optimization methods, particularly those trained with variational components is troubling.  It suggests that for sufficiently large problems, there may always be some distant defect which influences progress that could be made in the relevant region, simply due to chance.  We conjecture that this may be the physical mechanism at the heart of recent results in concentration of measure for the quantum approximate optimization method when examined on large enough instances~\cite{brandao2018fixed}.

As the shadow defects have their most direct impact in the outer loop optimization portion of short-depth variational optimizations, one solution is to modify the variational objective function from the expected value of the cost function to one which ignores higher energy contributions.  Two such suggestions have recently entered the literature, which are the conditional value at risk (CVaR)~\cite{barkoutsos2019improving} and a Gibbs-like function~\cite{li2019quantum} based on a log-sum-exp construction.  While CVaR is a reasonable choice, the function tends to be flat when looking at exact problems where the solution fraction is large due to its definition.  For this reason, we focus on the Gibbs inspired variational objective function
\begin{align}
    C_G(f(z)) = -\log \langle e^{-\eta f(z)} \rangle
\end{align}
where the brackets indicate the expected value over the current wavefunction.  Due to the diagonal nature of the cost function, this is simple to implement, and unlike the conditional value at risk, it maintains a non-flat surface even in the case where one can attain exact solutions.  We utilize an inverse temperature value of $\eta=20$ in our numerical experiments, as was done in the original paper.

We collect the results of the impact of both the cost function modification and the coherent cutting of the Laplacian in Fig. \ref{fig:ShadowDefectWithCut}.  We see that both the cost function modification and the Laplacian cutting can yield significant improvements.  While the cost function modification is easier to implement and has no additional quantum requirements, it avoids shadow defects at the cost of losing some fraction of the wavefunction under true higher potential regions.  Doing this for too many steps is ultimately not scalable, and ultimately one must find a way to constrain the wavefunction flow to relevant regions of space.  A coherent graph cutting method along with initialization in the relevant region is a first step in this direction, but requires additional quantum resources.  We now move to describing this method.

\section{Coherent graph cutting} \label{sec:CoherentCutting}
The study of the impact of kinetic energy as well as the inherent impact of shadow defects from the influence of the global surface suggests that modifying the support of the initial wavefunction and underlying graph can lead to significant improvements.  To maximize our capabilities in low depth, we have thus far focused on graph Laplacians that are amenable to fast-forwarding through diagonalization by a Hadamard transformation.  These Laplacians, often constructed in a Pauli framework by sums over $X$-like terms construct $k$-regular graphs in bitstring space, where $k$ is the number of terms in the sum.  It is easy to see that adding more of this type of term strictly increases the connectivity of the underlying graph, and in the limit of full connectivity, we recover the complete graph of Grover's algorithm, which we have seen cannot take advantage of any structure in the cost function.

In order to remove connections from a graph within a Pauli framework, it becomes necessary to introduce $XZ$ or $YY$ type terms.  These types of terms are sometimes broadly grouped into the class of non-stoquastic terms, but in this context if one used them to build a graph Laplacian that simply removed links built by $XX$ terms, the overall $L_G$ can remain stoquastic, allowing guarantees about adiabatic algorithms starting from the $\ket{+}^{\otimes n}$ state to remain true.  For the most general non-stoquastic deformation of $L_G$ it can be as challenging to determine the ground state of this $L_G$ to start from as to solve the original problem, whereas when the overall operator is guaranteed stoquastic, the ground state is always $\ket{+}^{\otimes n}$ independent of the non-stoquastic nature of individual terms.  Interestingly, it has recently been observed in the context of adiabatic optimization that non-stoquastic Hamiltonians can always be deformed to stoquastic Hamiltonians with larger gaps~\cite{crosson2020signing}, lending support to the strategy of considering these problems in the context of graph Laplacians that are always stoquastic as a whole in spite of individual terms.

The more dire consequences for our purposes of attempting to modify the graph structure with only Pauli operators is firstly that even conceptually simple modifications (e.g. cut around a given Hamming ball) seem to require a huge (exponential) number of terms and second that the addition of even a single $XZ$ or $YY$ term can ruin our ability to fast-forward.  We note that Trotter errors induced for the purposes of heuristic low-depth optimization may not actually be consequential, but we construct here a coherent graph modification that circumvents both of these problems.  In particular, we will use the ability of quantum operations to be efficiently controlled by functions with simple definitions, which do not always translate to simple (or polynomially sized) Pauli expansions.  

In short, we will take the quantum operation that enacts $e^{-i \beta \bar L_G}$ for a fast-forwardable LG $L_G$, and control its application on an ancilla indicating if vertices on the bit string graph that are within a certain domain.  When coupled with state initialization within that domain, this restricts the evolution to that domain without introducing excess kinetic energy from boundaries.

We consider domains that are defined by a Hamming ball of radius $r$ centered at a bitstring $\ket{x_0}$.  An efficient algorithm for preparing an equal superposition over this region, or approximates thereof has previously been given by Childs et al. using biased Hadamard transforms~\cite{childs2000finding}.  From this initial state, our restricted walk operator is formed as follows.  The state vector $\ket{\psi}$ is stored on $n$ qubits.  We attach an ancilla register of $O(\log n )$ qubits that stores the Hamming distance from the target central bitstring $\ket{x_0}$.  The Hamming distance from $\ket{x_0}$ is coherently computed on $\ket{\psi}$ and stored in this ancilla register.  A final indicator register with a single qubit is then used to store the result of a comparison function that computes if the value was less than the desired radius $r$.  The operator $e^{-i \beta \bar L_G}$ is performed, controlled on this indicator register.  This is followed by un-computation of the indicator and Hamming distance register.  This completes the application of a coherent, graph-cut evolution.

The results in Fig.~\ref{fig:ShadowDefectWithCut} show that this technique is more efficient at the elimination of shadow defects when compared to simple variational objective modification.  However the overhead of performing these coherent operations may make them prohibitive in the near-term.  While we focused on the simple case of Hamming ball restriction, this method can be generalized to more interesting functions of domain restriction and novel ways of specifying graphs.  Our results on kinetic energy and shadow defects indicate that developments in this area more generally will be needed to maximize the efficiency of quantum optimization methods.

\section{Improving low-depth approaches} \label{sec:ImprovingLowDepth}
Throughout this work we have collected insights about low depth quantum optimizations for classical problems in the hopes that they lead to simple and powerful new methods or improvements to existing methods.  Here we briefly review these ideas and assemble them into a framework that can be used to improve the potential for low-depth quantum optimization for classical problems.

One of the early challenges we identified was the vulnerability of quantum optimization methods to problems of scaling.  That is, even for uncoupled spins, if the energy scales of the spins differ appreciably, it can become needlessly challenging to obtain a solution, in analogy to condition number problems with first order classical methods.  If we insist that our method implements some quantum query to the objective function coherently, this leaves us with a logical building block of the form
\begin{align}
    e^{-i \gamma F(f(z); q)}
\end{align}
where $F$ is a general transformation of the objective with some schedule parameter $q$ such that there exists a $q$ where $F(f(z);q)=f(z)$.  This constraint allows us to define paths with this building block that coincide in a large depth limit to an adiabatic solution and hence coarser and coarser approximations of the exact limit can suggest logical building blocks for low-depth optimization circuits.  

However, if one is able to afford additional variational parameters, it is possible to define parameterized transformations or to simply relax existing approaches like QAOA by allowing more freedom.  Such a transformation allows one to solve the problem of non-interacting spins in constant depth by construction, however we note that this can pass the problem of scales into a classical outer loop optimizer.  Luckily second order optimizers or other preconditioning methods that can repair scale problems classically are well developed, unlike their quantum counterparts.  Hence one can take advantage of this additional classical technology to improve the efficiency of the optimization.  If one can heuristically guess the value of the preconditioners classically, this is obviously preferable.  Even with the efficiency of classical methods assisting variational optimization, these methods are believed to be limited by the ability to readout the relevant quantum information about the gradient, with scalings in the best case as $O(1/\epsilon)$.  Hence if the gap is exponentially small, it still may be difficult to resolve exactly without other techniques.  We examine numerically the case of varying $\gamma$ and $\beta$ independently for each term within QAOA in the subsequent section.

From connections to classical approximation algorithms, we noted that many of the quantum algorithms for optimization can be viewed as versions of continuous extensions followed by rounding through quantum measurements.  In this spirit, some of the problems we have reviewed and more can be simply eliminated without increasing quantum resources through the use of iterated rounding, where the number of frozen variables and other hyperparameters may be tuned.  In the next section we will examine the efficacy of this approach for several systems numerically.

Moving towards the dynamics of the system, we noted that the relationship between the graph Laplacian $L_G$, and the potential $f(z)$ was a crucial determiner of the amount of progress that could be made in each step.  We restricted ourselves to graphs that were fast-forwardable, including a generalization of these graphs that allowed coherent cutting of the domains for improved efficiency.  We noted that all graph Laplacians satisfy the requirement of allowing an exact path, hence there are a large class of transformations $K(L_G; q)$ that admit exact results, and indeed we have seen that even those outside the strict definition of a Laplacian are also applicable.  In a variational framework, it's clear that one may examine families of Laplacians and attempt to find correlations between a problem class over a defined measure of instances and the optimal Laplacians.  

When one uses fast-forwardable Laplacians as we have here, it implies the existence of an efficient diagonalizing quantum circuit.  In our case, we looked primarily at graph Laplacians that are diagonalized by a Hadamard transform across all qubits.  This means that measuring the kinetic energy is about as efficient as measuring the diagonal potential for these constructions.  That is, one nice feature of QAOA is that a single shot measurement of all qubits in the $Z$ basis allows complete determination of the energy, and the covariance between terms is often beneficial in the sense that estimating the energy to a given precision is relatively rapid compared to other cases. The construction of these estimators has been the topic of several recent works emphasizing this fact~\cite{huggins2019efficient,izmaylov2019unitary,gokhale2019n}. This diagonal form was also what allowed the use of simple modifications of the objective function as $C(f(z))$.  We see for our Hamiltonians this extends to the Laplacian in that one may measure all qubits in the $X$ basis, and infer not only an average of the Laplacian, but generalized functions of the Laplacian $K(L_G)$ that may be computed simply in its diagonal basis.  This may be simply summarized as
\begin{align}
    & \text{ measure } Z_i \forall \ i \rightarrow f(z) \rightarrow C(f(z)) \\
      H^{\otimes n} \rightarrow & \text{ measure } Z_i \forall \ i \rightarrow L_G(x) \rightarrow K(L_G)
\end{align}
where each is done some fraction of the time depending on statistics of convergence in the output results, and the diagonalizing circuit $H^{\otimes n}$ could be exchanged with another, for example a quantum Fourier transform on a Euclidean graph, or an XY graph in a fermionic picture~\cite{wang2018quantum,wang2020,chapman2020characterization}.  The ability to easily measure the Laplacian in its diagonal basis is convenient, as when one searches for the optimal transformation or Laplacian variationally with the intent to use it more generally, it is a mistake to look only at the decrease in the cost function expectation.  In particular, we saw through intuition and example that neglecting the effect of a particular primitive ansatz component on the kinetic energy resulted in less future information gain.

\begin{figure*}[t!]
\centering
\includegraphics[width=5.5in]{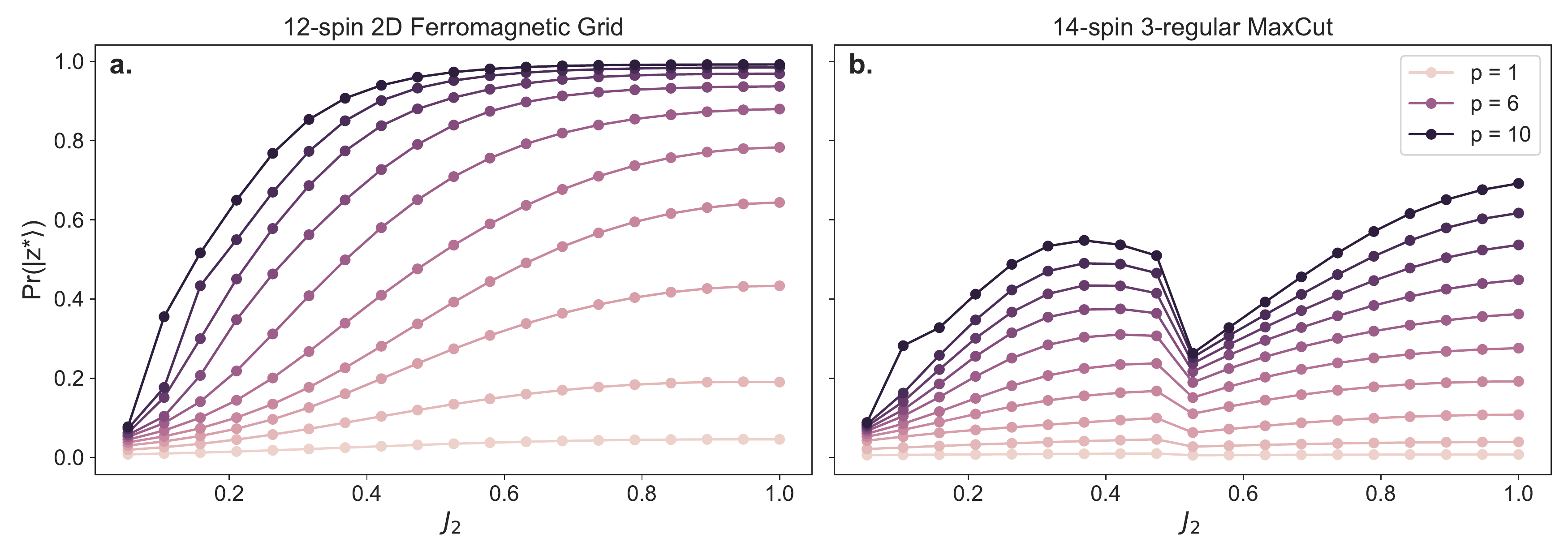}
    \caption{QAOA dependence on scale and detuning.  The 2D ferromagnetic ising model and 3-regular MaxCut problems are considered with two distinct couplings, and the dramatic performance changes are shown as the second coupling $J_2$ is altered.  Separate lines label different values of repetitions ($p$) for the QAOA ansatz.  \textbf{a.} The probability of the exact ground state of a 2D ferromagnetic chain decreases as half of the couplings are detuned. Increasing the number of QAOA rounds, $p$, delays the onset of such problems of scale, but performance remains strongly dependent on the separation.  \textbf{b.} A MaxCut-like problem instance that shows non-monotonic dependence on scale difference, with a curious dip at $J_2 \sim 0.52$ where a particular scaling of the couplings apparently substantially decreases performance of the algorithm.
    \label{fig:Numerics-Scale}}
\end{figure*}

From the study of the conflicted pair potential, we noted that one striking failure of some of these low depth optimization methods was when the averaged one-body potentials are in disagreement with the ground state.  One way to rectify this was to use freedom in the graph Laplacian to set for an individual qubit a value of $\beta_i = 3 \pi / 4$ rather than the value $\beta_i = \pi/4$.  Of course, it is key to do this on a per-qubit basis to retain the most flexibility.  Of course, one does not wish to introduce a combinatorial search over individual $\beta_i$ parameters, so a good heuristic strategy, such as greedy search over $\beta_i \in \{\pi/4, 3\pi /4\}$ for each qubit can provide a cheap alternative that we have seen can improve very low depth solutions dramatically.  We note that this bears a similarity to the suggestion of alternating $\beta_i$ stochastically in adiabatic optimizations to avoid gap closures with some finite probability~\cite{farhi2009quantum}.  This can be coupled with a continuous scaling by a more global $\beta$ value to not lose the flexibility of a QAOA result.  A flexible version of this strategy where $\beta$ is allowed to vary for each qubit variationally is examined later. 

When one is concerned with finding reusable primitives through variational pre-training, we advise the use of a modified hyper-cost function for this process that is coupled with the intended graph Laplacian.  That is, one should study re-usable components with a variational objective of the form
\begin{align}
    V(f(z), L_G) = k_1 C(f(z)) + k_2 K(L_G(x)) 
\end{align}
where the variational cost function transformation $C(f(z))$ includes generalizations like the Gibbs cost function and hyper-parameters $k_1$ and $k_2$ are to be adjusted.  Using a training cost function for individual components such as this allows one to quantify both gain in cost function, and anticipated information gain in future steps, which will ultimately be more scalable than relying on the cost function alone.

The use of the generalized cost functions $C(f(z))$ in any total variational methods allow one to mitigate the impact of shadow defects problems.  However, these approaches eventually lose efficiency as more and more of the wavefunction is discarded.  An efficient method must be able to restrict the domain of exploration intelligently.  We saw that if one remains entirely in a Pauli framework for $L_G$, this implies a loss of the ability to fast-forward, however the introduction of coherent cutting circumvents this problem.  This suggests that in a variational approach, the domain restriction could also be parameterized to great effect.  We do not explore this further here however.

Finally, we note that from connections with classical continuous extensions, specifically oblivious extensions, it need not be the case that a quantum coherent query is required to achieve an advantage.  This is especially appealing in the near-term, where the most expensive part of these applications is often the coherent query to the objective function, suggesting more problem specific and VQE like approaches may prove beneficial. Put together, these building blocks represent a variational framework for optimization built from concrete components informed by our understanding of the underlying mechanisms.  We believe further insights into the mechanisms will allow for additional improvements and more effective algorithms requiring even fewer quantum resources.

\section{Numerical Experiments} \label{sec:Numerics}

In this section, we examine some of the topics discussed throughout numerically for models that do not have simple analytical solutions.  This allows us to reason about the impact of these effects and techniques for more interesting systems, yet still simple enough to impart intuition for how these effects arise.  The impact of problems of scale are evaluated as well as two potential techniques for resolution, parameter relaxation and iterated rounding.

As described in Sec~\ref{sec:ScaleProblems}, a simple 1D ferromagnetic chain with non-uniform couplings can lead to an exponential gap closure with system size in quantum annealing~\cite{fisher1995critical}, which raises the natural question of how this trouble impacts methods like VQE and QAOA given the problems of scale discussed for non-interacting systems. To examine this, we first introduce a variant of this problem by modulating the alternative coupling coefficient $J_2$ resulting in the Hamiltonian
\begin{align}
H = \sum_{i=1}^{n/2} -Z_i Z_{i+1} + \sum_{i = n/2+1}^{n-1} -J_2 Z_i Z_{i+1}
\end{align}
as well as a two-dimensional variant where coupled spins are on a grid and the separate coupling values are given to different blocks. We performed numerical optimization of the QAOA ansatz for these problems and found that as the two domains of the problem are detuned---i.e. $J_2$ is scaled away from $1$---the performance of the QAOA as measured by resulting probability of the ground-state bitstrings $|z^*\rangle$ decreases analogous to the behavior noted in annealing. Performance on the 1D ferromagnetic chain decreases at the same rate regardless of the number of rounds $p$. The performance on the 2D ferromagnetic chain is depicted in Fig.~\ref{fig:Numerics-Scale}a and in addition to the overall downward trend as $J_2$ is decreased, higher values of $p$ can maintain higher probability mass on the ground state solutions for longer. 

We also consider a problem without a trivial solution, which is a modified version of  MaxCut given by the Hamiltonian
\begin{align}
H &= \sum_{i, j \in E} J Z_i Z_j \\
J &= \{1.0 \mathrm{\ or\ } J_2\}
\end{align}
In this example, the edges $E$ were defined by random 3-regular graphs and $J$ was selected such that half of the edges received a coupling strength of $J_2$. Unlike the simple ferromagnetic case, the behavior of these problems is more varied. Fig.~\ref{fig:Numerics-Scale}b shows a representative instance as a function of $J_2$. Across this and other random instances of this MaxCut problem, we observe a `spectrum' of performance with particular values of the coupling strength being relatively detrimental to the performance of the algorithm. Future research may lend insight into mechanisms underpinning this behavior.

\begin{figure}[htbp]
\centering
\includegraphics[width=0.40\textwidth]{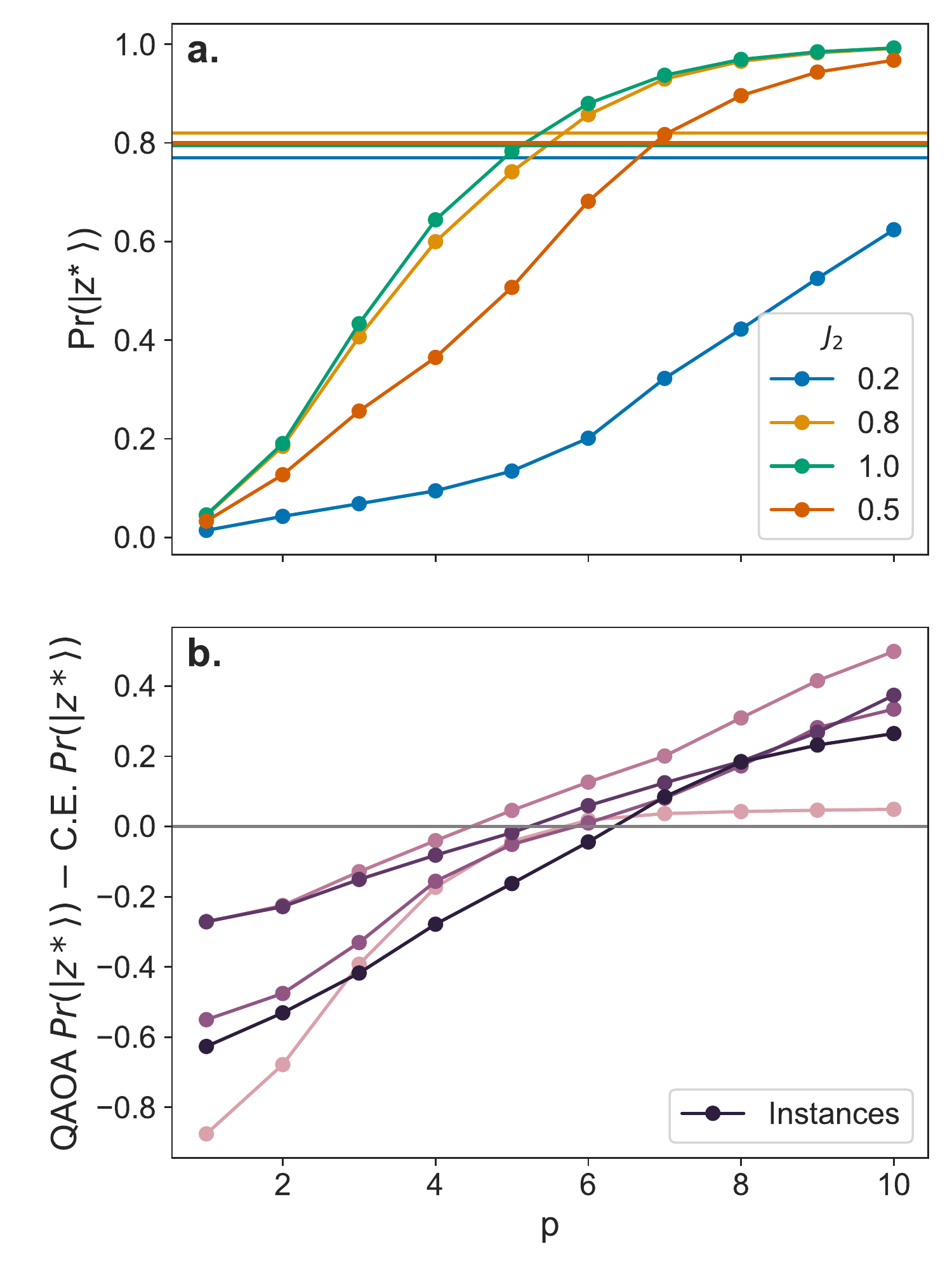}
    \caption{QAOA success probability against classical continuous extension baseline in problems of scale.  The same instances from Fig.~\ref{fig:Numerics-Scale} are used, with the horizontal lines indicating the performance of a classical continuous extension, and different lines for different detunings of the algorithm. \textbf{a.} For $n=12$, 2D ferromagnetic grids, we see that one may need values of $p$ in excess of 10 to match the classical baselines.  
    \textbf{b.} For $n=14$, 3-regular MaxCut problems ($J=1$), we see for a range of instances, one often needs a $p \geq 6$ to match the classical baseline.
    \label{fig:Numerics-CE}}
\end{figure}

To understand the value of using quantum queries to the classical Hamiltonian as in QAOA, classical continuous extensions in a VQE context can be used as a baseline of performance on these problems, as they are simple and do not demand a quantum query to the classical Hamiltonian. The ansatz here is a circuit of the form $\prod_i R_y(\theta_i)$, where the $\theta_i$ are optimized to minimize the expected value of the cost function $\langle C \rangle$, and we measure success once again by computing the probability of finding a correct bitstring $\text{Pr}(|z^*\rangle)$. Since this problem is not convex, we re-start the classical optimization many times from random parameter intializations to statistically estimate the overall probability of finding the ground state.  This gives a measure of how effective this simple ansatz is at convexifying the landscape.  We note that the optimization of this ansatz typically converges with all the support on one bitstring, so for a given restart there is either all or no support on a target bitstring, further motivating our approach to average over classical restarts. These probabilities can be directly compared to the probability of ground state bitstrings achieved through an optimized QAOA ansatz. The QAOA optimization is started from the point suggested by smooth annealing as has been suggested in previous studies, as random have been observed to produce inferior results~\cite{zhou2018quantum}.  This comparison is made in Fig.~\ref{fig:Numerics-CE}. Panel (a) treats the 2D ferromagnetic grid model. Following our previous discussion, the performance of QAOA is dependent on the $J_2$ parameter (lines with markers). The classical approach does not run into these problems of scale, with the continuous extension ansatz yielding $\sim 80\%$ probability of success regardless of $J_2$ value. QAOA without detuned domains ($J_2 \ge 0.8$) requires a number of rounds $p$ of approximately five to match the performance of the unentangled continuous extension. In contrast, for the highly detuned case ($J_2 = 0.2$) QAOA cannot match the performance of the classical algorithm even at $p = 10$. In panel (b) we once again turn to 3-regular MaxCut as representative realistic problems. Here we plot the difference in performance between the classical and quantum approaches (the greater variability of instances makes this a more digestible summary of results) with performance parity around $p \sim 6$. With many near-term hardware demonstrations of QAOA being restricted to $p < 5$~\cite{arute2020quantum}, this classical benchmark is all the more relevant.

\begin{figure}[t!]
\centering
\includegraphics[width=0.40\textwidth]{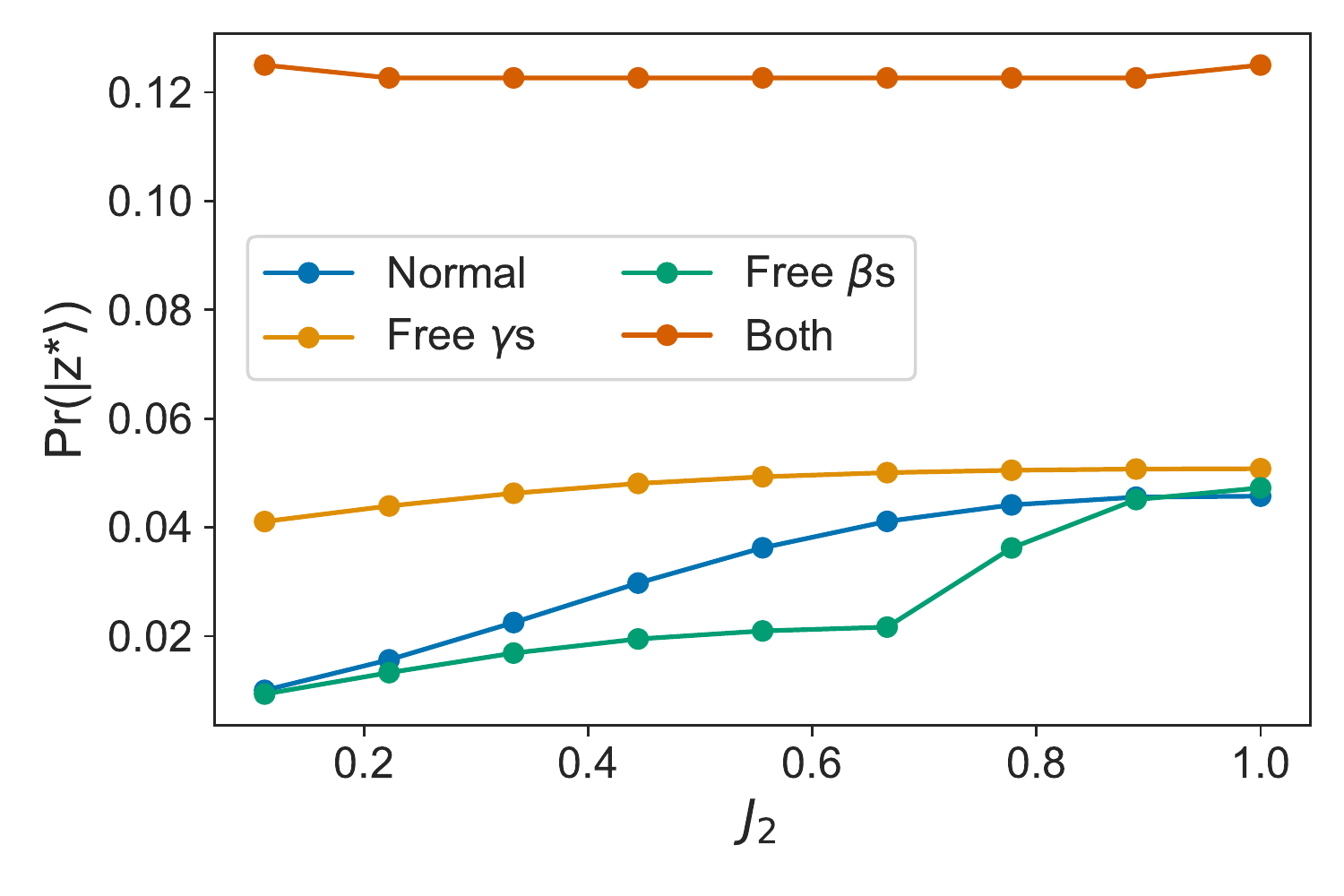}
\caption{Relative merit of QAOA relaxations in problems of scale for $p=1$.  Here we examine the effect for the same 2D problems of scale in relaxing for QAOA in turn the $\gamma$ and $\beta$ parameters.  As predicted from the non-interacting models, allowing free variation of $\gamma$ qualitatively restores the performance of QAOA across the variations in $J_2$, suggesting this is a simple but powerful modification.  Paradoxically, relaxation of only the $\beta$ terms leads to a decrease in performance under a probability of success metric, due to disagreement between the energy function and this objective under this constraint.  However we see that their combination gives a super-additive performance increase, at the cost of the greatest number of free parameters to optimize.
    \label{fig:Numerics-Freedom}}
\end{figure}

Having quantified the impact of problems of scale relative to a classical baseline, we turn to resolutions for this problem within QAOA.  In Fig~\ref{fig:Numerics-Freedom}, we consider relaxation in turn of the parameters $\gamma$, $\beta$, and their combination for a 2D ferromagnetic Ising model.  Following predictions from the non-interacting case, relaxation of the $\gamma$ parameters associated with the Hamiltonian afford the biggest qualitative change, leveling out and improving performance for more detuned values of $J_2$.  Similar to the intuition gained from seeing disagreement between lowering the energy and improving probability of success in previous sections on shadow defects, we see that simply varying the $\beta$ parameters independently actually reduces performance.  However their combination has an effect greater than the sum of the parts, improving performance across the board at the cost of having the most parameters to optimize.

\begin{figure}[t!]
\centering
\includegraphics[width=0.48\textwidth]{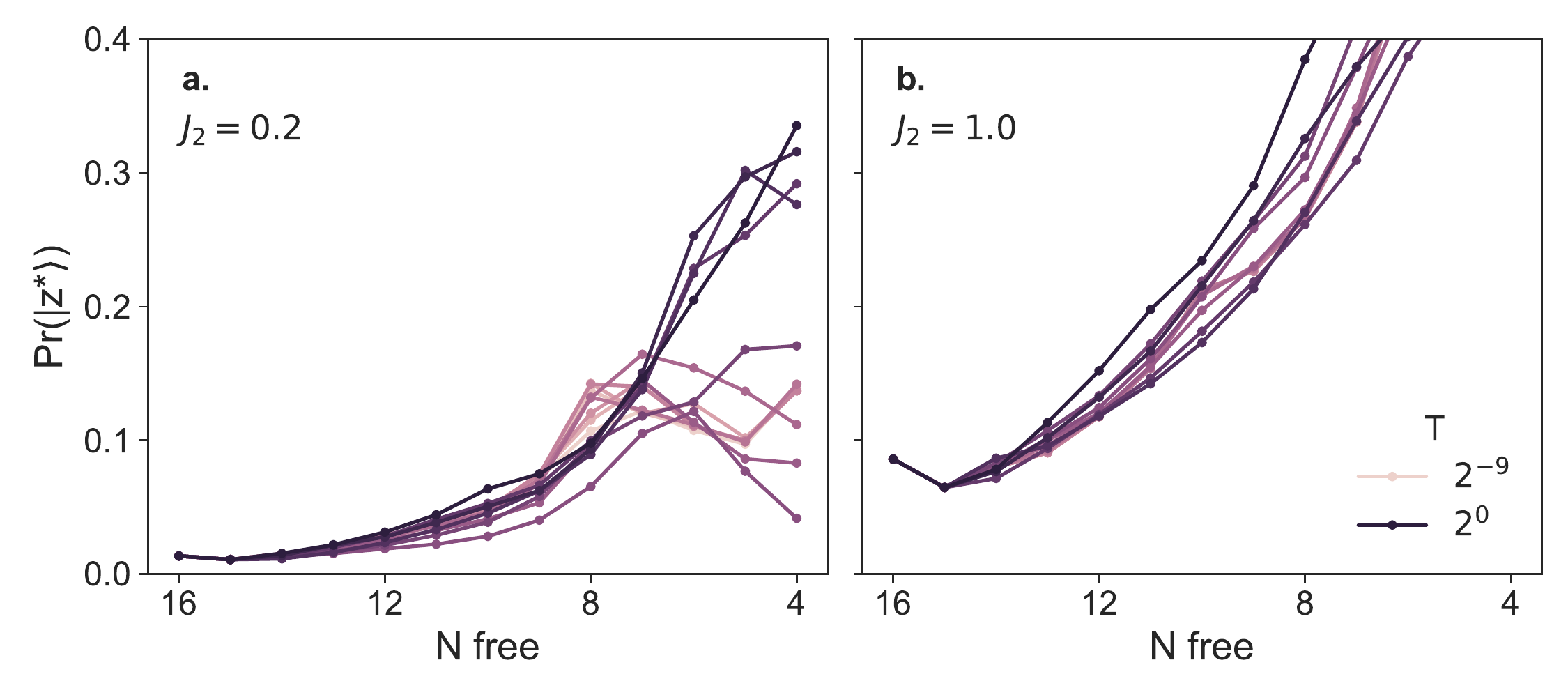}
\caption{Performance of iterated rounding technique for two settings of the 2D ferromagnetic.  
\textbf{a.} For the challenging $J_2=0.2$ case, as the number of variables are fixed, we see an increase in probability of success until a critical point around $6$ frozen variables where the performance bifurcates.  Even under bifurcation, the performance is strictly better than the unfrozen case. \textbf{b.} For simpler cases like $J=1$, iterated rounding is wildly successful in improving the quality of solution for a given value of $p$. \label{fig:Numerics-IterativeFixing}}
\end{figure}

We then examined the impact of iterated rounding for the instances of the 2D ferromagnetic model in Fig.~\ref{fig:Numerics-IterativeFixing}.  In both cases ($J_2=0.2$, $J_2=1$) we generally see an improvement in success as the variables are fixed.  However in the more challenging case, we see that there can be an optimal number of parameters to fix to improve success.  In the simpler case of $J_2=1$, rounding dramatically improves chances of success as a function of the number of frozen variables.

\section{Conclusions and outlook}
The ubiquity and impact of potential advances in optimization will continue to drive development of quantum algorithms.  As in classical methods development, the improvement and innovation of new algorithms depends on our understanding of the foundational elements that contribute to success, and what problems those inherently allow us to solve.  In this work, we identified physical mechanisms and intuition for a wide class of optimization algorithms by dissecting the performance of a single step, in order to draw attention to counter-intuitive limitations, and potential remedies.

We did so by first taking a general point of view that considers evolutions under fast-forwardable graphs of bitstrings.  This general class encompasses (in a limiting sense), adiabatic evolution, quantum annealing, quantum walks, and heuristics like VQE or QAOA.  Moreover, connections to continuous extensions in classical optimization approaches allow us to conjecture the origin of potential advantages for using oblivious ansatz like VQE for generic optimization problems on quantum computers. By examining a single step at different points in these trajectories we were able to identify some characteristics of this style of optimization that lead to better understanding of performance and methods for solution.  In particular, we were able to characterize classes of potentials that are exactly solvable for a given graph Laplacian, $L_G$, and the implications of uniqueness around this result.  From there, we showed that without modification, even simple non-interacting problems can be outside the scope of a single step, owing to basic limitations in the flexibility of the algorithms.  Numerically, we saw these limitations extended to larger problems like 2D ferromagnets, but simple fixes could often alleviate these problems. The case of exactly solvable potentials were linked to the measure vote mechanism, which demonstrated in comparison with its mean-field counterpart, that while entanglement plays a role, it can be strictly detrimental without enhancement of the graph Laplacian.

In addition, the importance of the quantum kinetic energy and its role in optimization was explored in detail.  The consequences of boundary and phase scattering are seen to make improvements grind to a halt, linking to diminished information gain from the potential. 
We saw that boundary scattering was linked with the ability of distant energy defects to impact the ability to optimize.  With a large enough space and a wide enough variety of defects, we conjectured these are the underlying origin of concentration phenomena that have been observed in QAOA landscapes. These shadow defects can be partially circumvented through modification of the variational objective, but ultimately coherent cutting or clever modification of the graph Laplacian $L_G$ is required to improve efficiently.

Overall we believe that providing more physical intuition for the operation of algorithms for quantum optimization will open the doors to dramatic algorithm improvement.  Ultimately the lack of a globally convergent, state independent quantum optimization method implies that we will either eventually require state dependent methods like reverse annealing, or some degree of non-unitarity with respect to the optimization system using an ancillary system or measurement.  We believe the insights in this work will guide the development of improved future methods, free from some of the deficiencies we have outlined here, and bolstered by their remedies.

\section{Acknowledgements}
We thank Eleanor Rieffel for detailed comments and feedback on the draft.

\bibliographystyle{apsrev4-1_with_title}
\bibliography{references}

\end{document}